\documentclass[twocolumn,english,prb,twocolumn,floatfix,footinbib,superscriptaddress,showpacs, nokeys,notitlepage]{revtex4-1}
\usepackage[T1]{fontenc}
\usepackage[latin9]{inputenc}
\usepackage{color}
\usepackage{float}
\usepackage{amsmath}
\usepackage{amssymb}
\usepackage{graphicx}
\usepackage{esint}

\makeatletter

\providecommand{\tabularnewline}{\\}

\usepackage[usenames,dvipsnames,svgnames,table]{xcolor}
\usepackage[UKenglish]{babel}
 \usepackage[colorlinks]{hyperref}
\usepackage{hyphenat}

\usepackage[explicit]{titlesec}
\hypersetup{
     colorlinks   = true,
     citecolor    = blue,
     linkcolor=blue,
     urlcolor=blue
}

\makeatother

\usepackage{babel}
\begin{document}
\title{The Lipkin-Meshkov-Glick model with Markovian dissipation - A description
of a collective spin on a metallic surface}
\author{João S. Ferreira}
\email{joao.ferreira@unige.ch}

\affiliation{CeFEMA, Instituto Superior Técnico, Universidade de Lisboa Av. Rovisco
Pais, 1049-001 Lisboa, Portugal}
\author{Pedro Ribeiro}
\email{ribeiro.pedro@gmail.com}

\affiliation{CeFEMA, Instituto Superior Técnico, Universidade de Lisboa Av. Rovisco
Pais, 1049-001 Lisboa, Portugal}
\begin{abstract}
\noindent\begin{minipage}[t]{1\columnwidth}%
\global\long\def\ket#1{\left| #1\right\rangle }%

\global\long\def\bra#1{\left\langle #1 \right|}%

\global\long\def\kket#1{\ket{\ket{#1}}}%

\global\long\def\bbra#1{\bra{\bra{#1}}}%

\global\long\def\braket#1#2{\left\langle #1\right. \left| #2 \right\rangle }%

\global\long\def\bbrakket#1#2{\left\langle #1\right. \left\Vert #2\right\rangle }%

\global\long\def\av#1{\left\langle #1 \right\rangle }%

\global\long\def\tr{\text{tr}}%

\global\long\def\Tr{\text{Tr}}%

\global\long\def\pd{\partial}%

\global\long\def\im{\text{Im}}%

\global\long\def\re{\text{Re}}%

\global\long\def\sgn{\text{sgn}}%

\global\long\def\Det{\text{Det}}%

\global\long\def\abs#1{\left|#1\right|}%

\global\long\def\up{\uparrow}%

\global\long\def\down{\downarrow}%

\global\long\def\k{\mathbf{k}}%

\global\long\def\wks{\mathbf{\omega k}\sigma}%

\global\long\def\vc#1{\mathbf{#1}}%

\global\long\def\bs#1{\boldsymbol{#1}}%

\global\long\def\t#1{\text{#1}}%

\global\long\def\L{\mathcal{L}}%

\global\long\def\r{\mathcal{\rho}}%

\global\long\def\sz{S_{z}}%

\global\long\def\sx{S_{x}}%

\global\long\def\sy{S_{y}}%

\global\long\def\sp{S_{+}}%

\global\long\def\sm{S_{-}}%

\global\long\def\d{\dagger}%

\global\long\def\si{s\rightarrow\infty}%
\end{minipage}Motivated by recent prototypes of engineered atomic spin devices,
we study a fully connected system of $N$ spins $1/2$, modeled by
the Lipkin-Meshkov-Glick (LMG) model of a collective spin $s=N/2$
in the presence of Markovian dissipation processes. We determine and
classify the different phases of the dissipative LMG model with Markovian
dissipation, including the properties of the steady-state and the
dynamic behavior in the asymptotic long-time regime. Employing variational
methods and a systematic approach based on the Holstein-Primakoff
mapping, we determine the phase diagram and the spectral and steady-state
properties of the Liouvillian by studying both the infinite-$s$ limit
and $1/s$ corrections. Our approach reveals the existence of different
kinds of dynamical phases and phase transitions, multi-stability and
regions where the dynamics is recurrent. We provide a classification
of critical and non-critical Liouvillians according to their spectral
and steady-state properties.
\end{abstract}

\maketitle

\section{Introduction}

Quantum systems submitted to non-equilibrium conditions support a
rich set of physical phenomena yet to be classified. This endeavor
encompasses emergent features found in non-linear classical dynamics
and equilibrium quantum matter, but has also the potential to reveal
effects unique to non-equilibrium quantum degrees of freedom. Various
of these aspects have been explored recently, motivated by advances
in the manipulation and control of cold atomic and solid-state setups.

Artificial magnetic structures deposited on metallic surfaces are
particular examples of novel setups, where the ability to manipulate
and monitor individual atomic spins offers the possibility to study
a non-equilibrium quantum open system in a controlled fashion \citep{Heinrich2004,Hirjibehedin2007a,Tsukahara2009,Gauyacq2012}.
A number of prototypes have already demonstrated the potential of
these engineered atomic spin devices for information processing \citep{Leuenberger2001,Troiani2005,Imre2006a,Khajetoorians2011a,Baumann2015,Kalff2016}
and spintronics applications \citep{Heinrich2004,Hirjibehedin2006a,Loth2010,Loth2012a}.
The basic setup consists of a set of magnetic atoms deposited on a
thin insulating layer coating a metallic surface. Atoms are individually
addressable by a spin-polarized metallic tip. Applying a finite bias
voltage between the tip and the surfaces induces an inelastic current
that can be used to infer properties of the magnetic state \citep{Fernandez-Rossier2009,Lorente2009,Delgado2010a,Ternes2015,Shakirov2017,Shakirov2016}.
For artificial magnetic structures, the most relevant system-environment
interaction is the magnetic exchange with the itinerant electrons
of the metallic substrate \citep{Delgado2010a,Delgado2014}. The environment
induces an effective memory on the dynamics of the system's density
matrix. Although memory effects are generically non-negligible, they
can, in some cases, be assumed instantaneous as compared with time-scales
within the system. For metallic environments, this Markovian regime
is obtained for large temperatures or chemical potentials \citep{Ribeiro2014e}.
In this work, we consider regimes where the bias voltage applied between
the tip and the metallic substrate is large. In this case, the master
equation for the evolution of the density matrix of the magnetic system,
obtained in Ref. \citep{Shakirov2016}, is Markovian and reduces to
the Lindblad equation \citep{H.-P.BreuerandF.Petruccione2002,DeVega2017}.

We examine the case of a fully connected magnetic structure made of
$N$ spins-$1/2$ and study the dynamics in the highest spin sector,
which can be modeled by a collective spin $s=N/2$. In the absence
of dissipation, collective spin models have been extensively investigated.
Perhaps, one of the best studied is the Lipkin-Meshkov-Glick (LMG)
model \citep{Lipkin1965,Meshkov1965,Glick1965} - a ubiquitous system
featuring a fully connected set of spins-$1/2$. Its ground-state
properties \citep{Cirac1998,Garanin1998,Latorre2005,Vidal2004,Vidal2004b},
spectrum, correlation functions \citep{Ulyanov1992,Turbiner1988,Ribeiro2007,Ribeiro2008,Ribeiro2009}
and dynamics \citep{Vidal2004b,Dusuel2004,Dusuel2005,Das2006,Hamdouni2007}
can be systematically obtained in the thermodynamic limit, i.e. large
$s$ limit, by a semi-classical expansion with $1/s$ playing a role
similar to $\hbar$. Non-perturbative effects can also be captured
by semi-classical methods \citep{Ribeiro2009}.

Markovian dissipation in collective spin models was first considered
to describe spontaneous emission of an ensemble of two-level atoms
in a superradiant phase \citep{Kilin1978,Drummond1978,Drummond1980,Carmichael1999}.
Various variants and generalizations of these models have been studied
since then \citep{Schneider2002,Morrison2008,Kessler2012,Hannukainen2017,Iemini2017}.
These systems belong to a family that we refer to as dissipative Lipkin-Meshkov-Glick
models, in analogy with its dissipationless counterpart. In cases
where an exact construction of the steady-state exists \citep{Puri1979,Drummond1980}
correlation functions can be computed exactly. Otherwise, semi-classical
methods \citep{Schneider2002,Morrison2008} and perturbative $1/s$
expansions \citep{Kessler2012} were employed, as well as exact diagonalization,
to access the steady-state and the spectrum of the Lindblad operator.
Such studies revealed the existence of several phases characterized
by qualitatively different steady-states properties. These include
systems with a single or bistable steady-states \citep{Morrison2008}
or cases where, in the thermodynamic limit, no steady-state could
be found and the system attains a recurrent periodic orbit, dependent
on its initial condition \citep{Kilin1978,Drummond1978,Drummond1980,Carmichael1999}.
Recently, models featuring independent, i.e. non-collective, spin
decay have also been considered \citep{Lee2014a,Kirton2017,Shammah2017}.

Contrarily to their equilibrium counterparts, a classification of
quantum critical phenomena in the presence of dissipation has not
yet been accomplished despite the significant body of works devoted
to the topic \citep{Prosen2010a,Eisert2010,Znidaric2011a,Honing2012,Lesanovsky2013,Horstmann2013,Genway2014,Sieberer2015,Znidaric2015,Casteels2017,Hannukainen2017}.
In particular, dissipative phase transitions have been shown to escape
Landau's symmetry breaking paradigm \citep{Eisert2010,Honing2012,Hannukainen2017}
in some cases but not others \citep{Sieberer2015}.

In this paper, we propose a classification of the phases of collective
spin models with Markovian dissipation according to their steady-state
and spectral properties. To do so, we study the different phases of
the dissipative LMG model with Markovian dissipation. The specific
form of the jump operators is motivated by a solid-state setup, which
features magnetic atoms deposited on a metallic surface, and where
spin transport arises by the proximity with a spin-polarized metallic
tip held at a finite bias voltage (Fig. \ref{fig:sys}). To access
these properties of the model, we employ variational methods, a systematic
Holstein-Primakoff mapping and exact diagonalization studies of the
Liouvillian.

Besides helping to understand non-equilibrium states of engineered
solid-state devices, our results are also of interest to quantum optics
and cold atomic setups, where dissipative phase transitions in optical
cavities \citep{Rodriguez2017,Fitzpatrick2017,Letscher2017,Brennecke2013}
have been observed which can be modeled by variants of the dissipative
LMG model.

The paper is organized as follows. The model is introduced in Sec.
\ref{sec:Model-and-Phase}. A description of the phase diagrams obtained
for two tip polarization directions, as well as the main characteristics
of each phase and phase transitions, are given in Sec. \ref{sec:Phase-Diagram}.
Sec. \ref{subsec:Holstein-Primakoff-transformed-L} gives a summary
account of the $1/s$ expansion using the Holstein-Primakoff mapping
that can be used to systematically compute $1/s$ corrections of observables.
A detailed analysis of the Liouvillian spectrum, dynamics and properties
of the steady-state in each of the phases, as well as the phase transition
lines are given in Sec. \ref{sec:Linearized-Liouvillian-operator}.
In Sec. \ref{subsec:Classification-of-steady-state} we give a classification
of the different phases and summarize our main findings. We conclude
in Sec. \ref{sec:Discussion} with the implications of our work. The
Appendix sections present some of the details of calculations used
to derive the results in the main text. Sec. \ref{sec:Equations-of-Motion}
provides a derivation of the semi-classical and variational equations
of motion. Sec. \ref{sec:Dynamics-for-finite-s} contains helpful
simulations of the magnetization dynamics for finite-$s$ systems
and Sec. \ref{sec:The Linearized Lindblad Operator} details the derivation
of the linearized Liouvillian.

\section{Model\label{sec:Model-and-Phase}}

\begin{figure}
\centering{}\includegraphics[height=4cm]{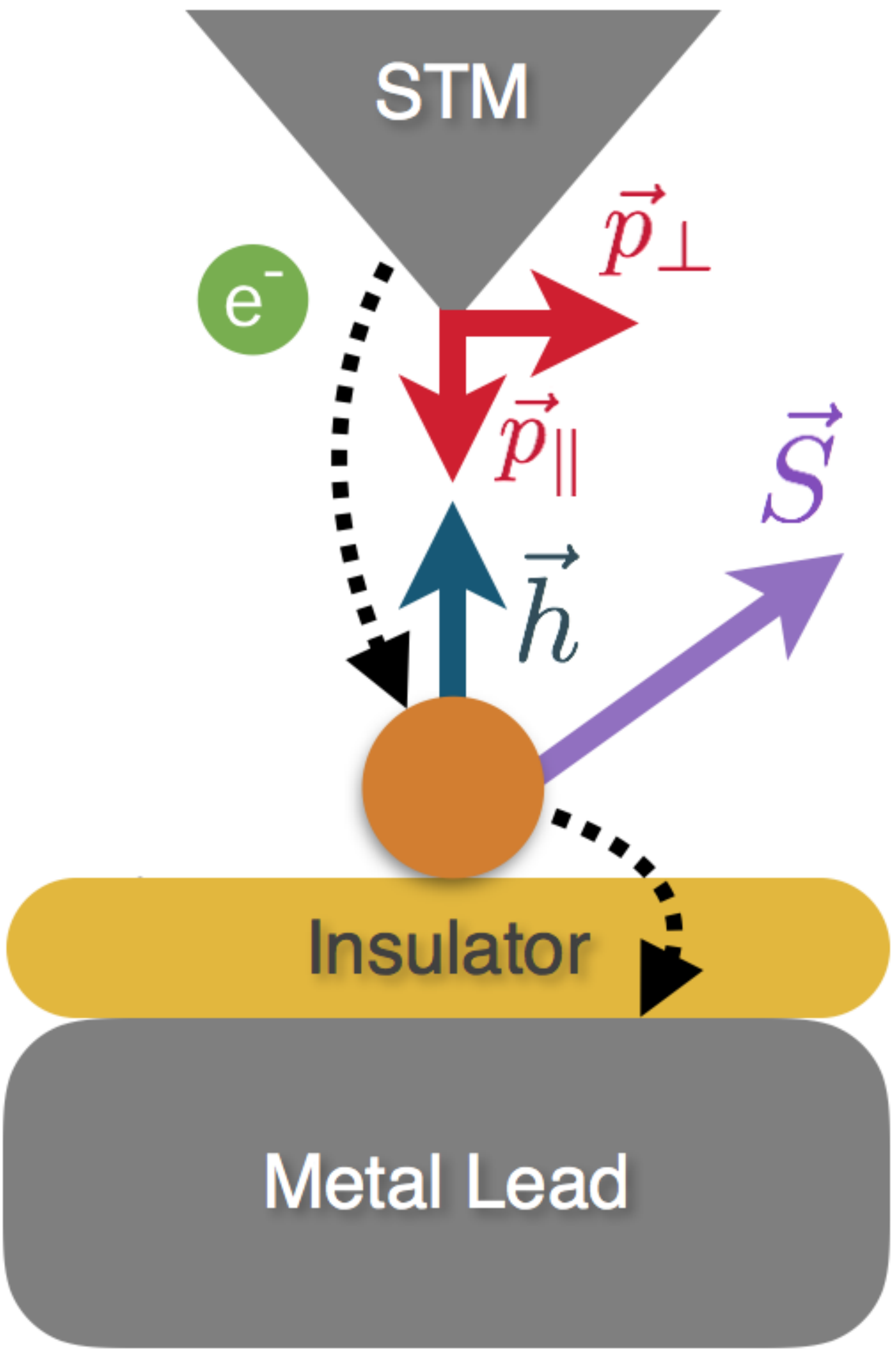}\caption{\label{fig:sys}(a) Schematic representation of the setup. A collective
moment is obtained as an effective description of an aggregate of
$N=2s$ magnetic atoms, with a large charge-gap, deposited on an insulating
layer coating a metallic substrate. Upon applying a voltage difference
between the metallic spin-polarized tip and the substrate, a charge
current ensues. Two polarization directions are considered: $\protect\bs p=p\protect\bs e_{z}$
(i.e. $\protect\bs p\parallel\protect\bs h$) and $\protect\bs p=p\protect\bs e_{y}$
(i.e. $\protect\bs p\perp\protect\bs h$).}
\end{figure}

We consider the system depicted in Fig. \ref{fig:sys}, consisting
of a magnetic moment deposited on a metallic surface and in contact
with a metallic tip having a spin polarization vector $\bs p$. The
collective magnetic moment that can be of an atom, an artificial atomic
structure or a molecule, is modeled by the LMG Hamiltonian 
\begin{equation}
H=-\bs h.\bs S-\frac{1}{2s}\left(\gamma_{x}S_{x}^{2}+\gamma_{y}S_{y}^{2}\right)\label{eq:hamiltonian-1}
\end{equation}
with $S_{\alpha=x,y,z}$ obeying the $su(2)$ commutation relations
with $\bs S.\bs S=s(s+1)$. This spin representation is obtained as
the symmetric sector of $N=2s$ two-level systems. The coefficients
$\gamma_{x},\gamma_{y}$ are determined by the surface anisotropy
and $\bs h$ is a local magnetic field. In what follows, we consider
that the applied field always points in the direction perpendicular
to the surface, i.e. $\bs h=h\bs e_{z}$, and two possible orientations
for the polarization vector: a case where the field and the polarization
are parallel, with $\bs p=p\bs e_{z}$; and a case where they are
perpendicular, with $\bs p=p\bs e_{y}$ with $-1\leq p\leq1$.

The collective magnetic moment is a good effective description of
an atomic aggregate with a large charge-gap. The exchange interaction
between the magnetic moment and electrons in the metallic leads is
induced by virtual processes where the atomic aggregate acquires (donates)
and donates (acquires) an electron from the leads. Such processes
induce relaxation and decoherence effects to the magnetic state and
allow a charge current to ensue in the presence of a finite applied
voltage. If the effective exchange coupling is not too strong, a perturbative
treatment allows for the description of the dynamics in terms of a
(non-Markovian) master equation for the density matrix of the magnetic
moment; the details of the derivation can be found in Ref. \citep{Shakirov2016}.

A simple limit is recovered for a large bias voltage, where the environment
becomes memoryless. In this limit, the effect of the leads is simply
to perform spin-flips at a constant rate. In case the leads are spin
polarized, this yields a net spin transfer. In this Markovian limit
the Liouvillian super-operator, $\L$, determining the evolution of
the system's reduced density matrix, $\pd_{t}\rho=\L\left(\rho\right)$,
acquires the Lindblad form \citep{H.-P.BreuerandF.Petruccione2002,DeVega2017}

\begin{equation}
\L\left(\rho\right)=-i[H,\rho]+\sum_{i}W_{i}\rho W_{i}^{\dagger}-\frac{1}{2}\left\{ \rho,W_{i}^{\dagger}W_{i}\right\} \label{eq:lindblad-1-1-1}
\end{equation}
where $W_{i}$, with $i=+,-,z$, are the so called jump operators
\begin{equation}
W_{z}=\sqrt{\frac{\Gamma}{2s}}\tilde{S}_{z};\:W_{+}=\sqrt{\frac{\Gamma}{2s}\frac{1-p}{2}}\tilde{S}_{+};\:W_{-}=\sqrt{\frac{\Gamma}{2s}\frac{1+p}{2}}\tilde{S}_{-}\label{eq:w_op_model-1}
\end{equation}
The tilde ``$\,\tilde{\,}\,$'' denotes that the quantization axis
of the operator is taken along the polarization of the tip. In the
two situations treated here, we have $\tilde{S}_{\alpha}=S_{\alpha}$
for the parallel case and $\tilde{S}_{\alpha}=e^{i\frac{\pi}{2}S_{x}}S_{\alpha}e^{-i\frac{\pi}{2}S_{x}}$
for the perpendicular setup. $\Gamma$ is the rate of the quantum
jumps, proportional to the absolute value of the applied voltage (see
Appendix G of Ref. \citep{Shakirov2016}).

Under Liouvillian dynamics, the evolution of the density matrix is
given by 
\begin{equation}
\rho(t)=e^{t\L}\rho(t_{0})=\sum_{i}\exp\left(t\Lambda_{i}\right)X_{i}\tr\left[\tilde{X}_{i}\rho(t_{0})\right]\label{eq:-1}
\end{equation}
where $X_{i}$ and $\tilde{X}_{i}$ are respectively left and right
eigenvectors of the super-operator corresponding to the eigenvalue
$\Lambda_{i}$ and normalized such that $\tr\left[X_{i}\tilde{X}_{j}\right]=\delta_{ij}$.
The real part of $\Lambda_{i}$ is non-positive and there is at least
one zero eigenvalue $\Lambda_{0}=0$ corresponding to left eigenvector
$\tilde{X}_{0}=1$.

\begin{figure}[H]
\includegraphics[width=1\columnwidth]{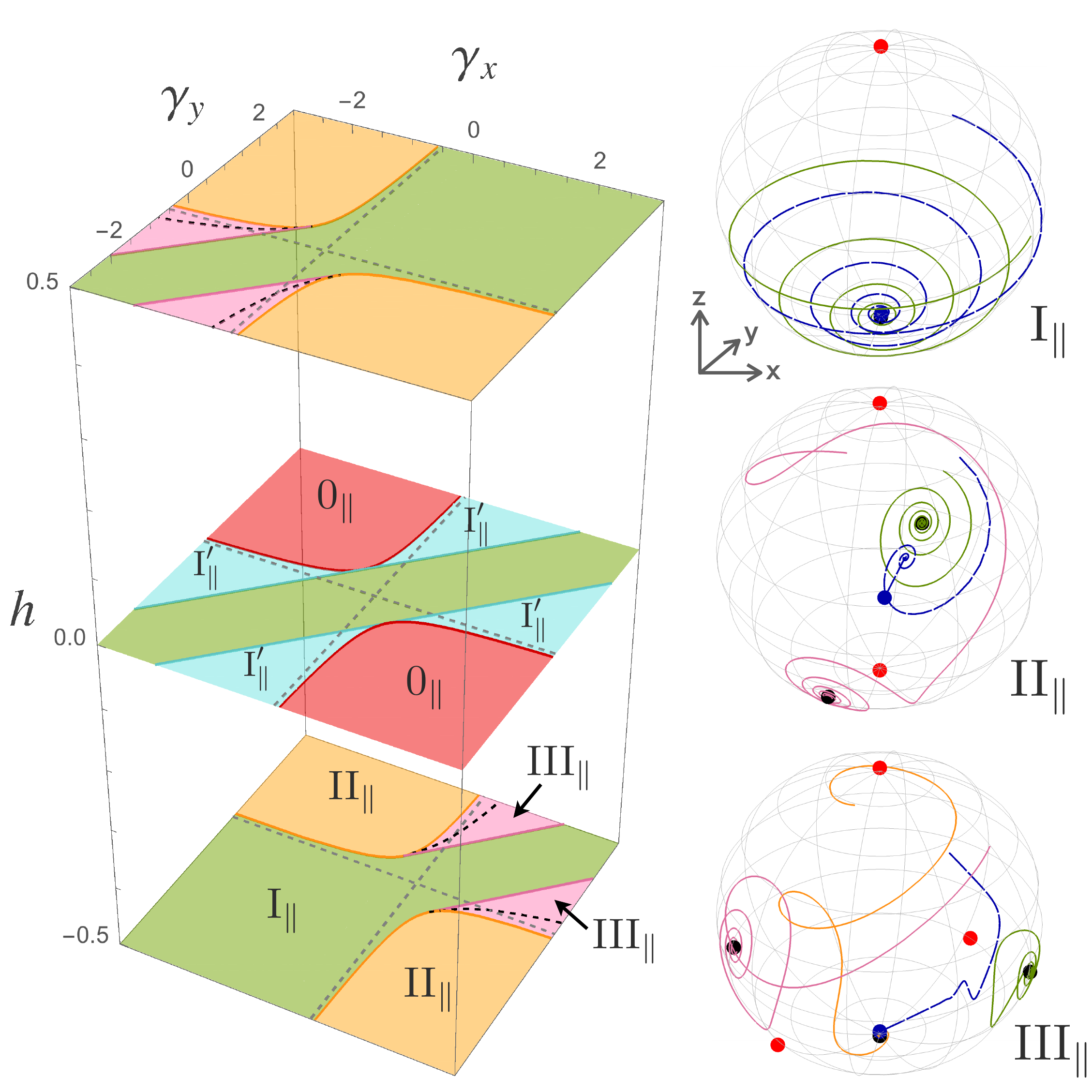}

\caption{\label{fig:pd_pz}Left panel: Phase diagram for $\protect\bs p\parallel\protect\bs h$,
plotted for $p\Gamma=1$. Right panels: Average magnetization $\protect\bs n=\protect\av{\protect\bs S}/s$
shown for representative states of each phase. Stable (unstable) infinite-$s$
steady-states are depicted as black (red) points. Representative trajectories
in the $s=\infty$ limit are represented in full colored (pink, green
and orange) lines. The steady state and the evolution of the magnetization
for $s=20$ are depicted by a blue point and blue dashed line respectively.
Parameters: $\protect\t I_{\parallel}$ - $h=1,\gamma_{x}=-0.2,\gamma_{y}=-0.3,p\Gamma=0.2$;
$\protect\t{II}_{\parallel}$ - $h=1,\gamma_{x}=0.5,\gamma_{y}=-2.5,p\Gamma=1$;
$\protect\t{III}_{\parallel}$ - $h=1,\gamma_{x}=-3,\gamma_{y}=-1.5,p\Gamma=1$.}
\end{figure}

\begin{figure}
\includegraphics[width=1\columnwidth]{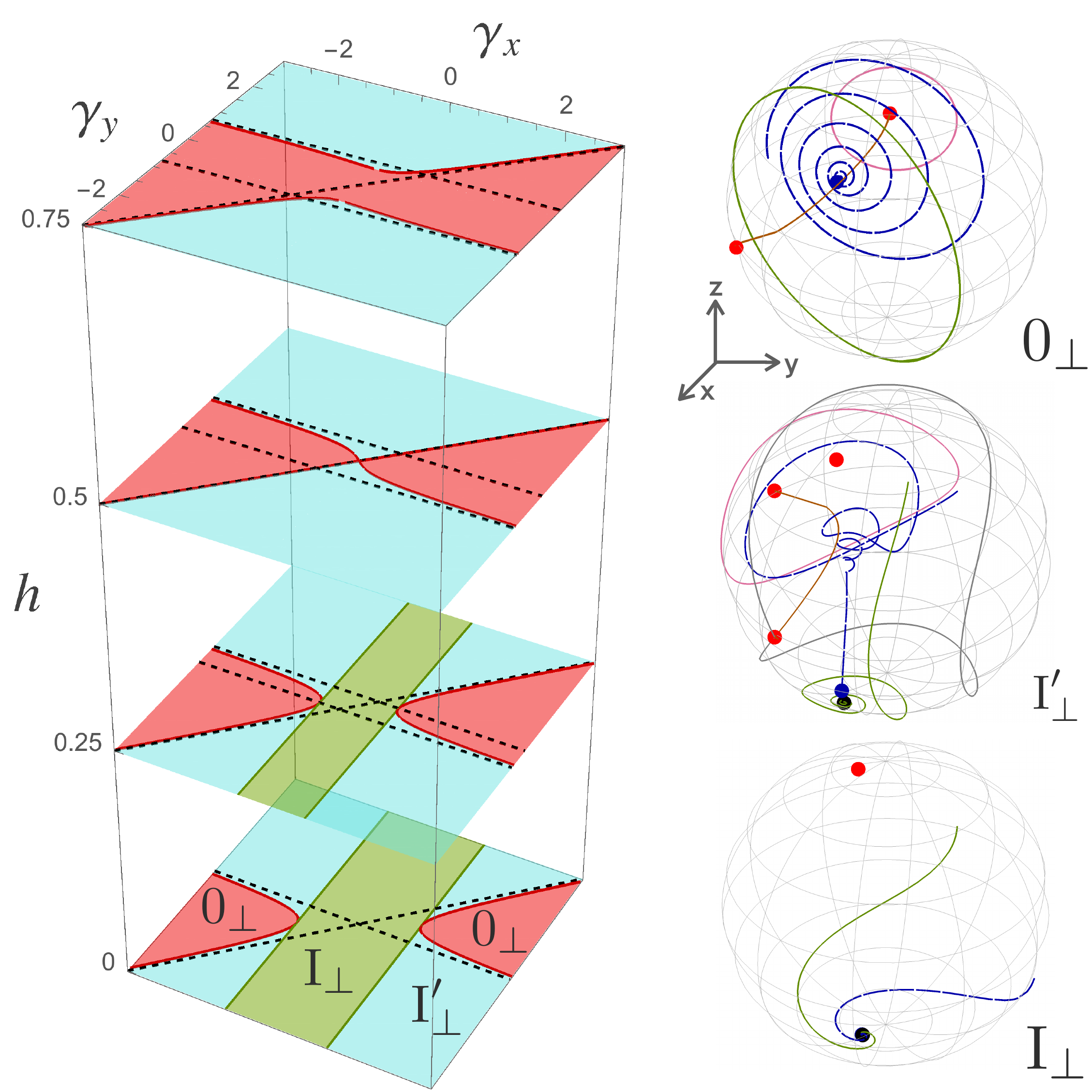}

\caption{\label{fig:pd_py}Left panel: Phase diagram for $\protect\bs p\perp\protect\bs h$.
Right panel: $\protect\bs n=\protect\av{\protect\bs S}/s$ shown for
states for the different phases. The color codes are the same as in
Fig.\ref{fig:pd_pz}. In the middle panel ($\protect\t I'_{\perp}$),
the separatrix separating the two qualitative long-time states of
regions $\text{0}_{\perp}$ and $\protect\bs{\text{I}}_{\perp}$ is
depicted in gray. The plots are done for the following set of parameters:
$\text{0}_{\perp}$- $h=1,\gamma_{x}=0.1,\gamma_{y}=0.2,p\Gamma=1$;$\text{I}_{\perp}$-
$h=1,\gamma_{x}=-2,\gamma_{y}=2.1,p\Gamma=1$;$\text{I'}_{\perp}$-
$h=0.2,\gamma_{x}=0.1,\gamma_{y}=1,p\Gamma=1$.}
\end{figure}

\section{Steady-state Phase Diagram\label{sec:Phase-Diagram}}

In this section, we determine the phase diagram of the model and characterize
the different phases according to the qualitative properties of the
steady-states. As in equilibrium, non-analyticities in the steady-state
observables are only expected once the thermodynamic limit is taken,
i.e $N=2s\to\infty$. Since, within the symmetric sector, the total
angular momentum is determined by $s=N/2$, the thermodynamic limit
corresponds to that of a large classic spin, $s\to\infty$.

To approximate the dynamics of $\rho\left(t\right)$ in the large
$s$ limit, we assume an ansatz density matrix of the form $\rho\propto e^{\bs m.\bs S}$,
and derive the equation of motion for the vector $\bs m$. Away from
phase transition points, this ansatz becomes exact in the $s\to\infty$
limit and allows for higher order corrections in powers of $1/s$.
In Appendices \ref{sec:Variational-density-matrix} and \ref{sec: Semiclassical dynamics},
we provide the details of the method and show how this approach compares
with the standard mean-field approximation \citep{Kilin1978,Drummond1978,Drummond1980,Carmichael1999}.

From the ansatz parameter $\bs m$, we compute the rescaled magnetization
vector $\bs n=\av{\bs S}/s$ and solve the fixed-point condition $\pd_{t}\bs n=0$
in order to obtain the steady-state magnetization. The fixed-points
of the dynamics are classified as attracting (stable), repulsive (unstable),
mixed (saddle-points, having at least one attractive and one repulsive
direction) or marginal (no attractive or repulsive direction), according
to the dynamics in their vicinity. Regarding steady-state properties,
different phases are characterized by the number and nature of the
fixed-points. A change in the number or nature of the fixed points
typically corresponds to non-analyticities of certain observables
as well as in the slowest decaying rate towards these points.

We recall that, while all fully-polarized vectors, i.e. $\abs{\bs n}=1$,
correspond to pure states, vectors with $\abs{\bs n}<1$ may correspond
both to pure or mixed states.

In the following two sub-sections, we study the two cases shown in
Fig. \ref{fig:pd_pz} and \ref{fig:pd_py}, corresponding to an applied
field parallel ($\bs p\parallel\bs h$) or perpendicular ($\bs p\perp\bs h$)
to the polarization. We qualitatively describe the different phases,
as well as the nature of the phase transitions between them based
on the steady-state properties and dynamics. The spectral analysis
within each phase is relegated to Sec. \ref{sec:Linearized-Liouvillian-operator}.

The main findings of this section are summarized in Sec. \ref{subsec:Classification-of-steady-state}
(see Table \ref{tab:Classification-of-steady-state}).

\subsection{Parallel polarization\label{subsec:Parallel-polarization-III}}

For parallel polarization ($\bs p\parallel\bs h$) {[}see Fig. \ref{fig:pd_pz}-(left
panel){]}, there are three stable phases, $\t I_{\parallel}$, $\t{II}_{\parallel}$
and $\t{III}_{\parallel}$, in the $\gamma_{x}-\gamma_{y}-h$ parameter
space, separated by critical surfaces where phase transitions occur.
Regions $\t 0_{\parallel}$ and $\t I'_{\parallel}$, arising only
at $h=0$, are also critical and correspond either to $\t I_{\parallel}\leftrightarrow\t{III}_{\parallel}$
transitions or to transitions between phases $\t{II}_{\parallel}$
with different steady-state symmetries. The critical phases $\t 0_{\parallel}$
and $\t I'_{\parallel}$ are similar to some of the phases found in
the perpendicular case ($\bs p\perp\bs h$) and we relegate their
study for the next subsection. While phases $\t I_{\parallel}$, $\t{II}_{\parallel}$
and $\t{III}_{\parallel}$ can be distinguished by their number of
fixed points (1,2 and 3), the further division within region $\t{III}_{\parallel}$,
depicted as a dashed black line, is obtained by considering steady-state
properties at finite-$s$ (see below).

Fig. \ref{fig:pd_pz}-(right panels) illustrates the dynamics of the
average magnetization, $\av{\bs S}$, within each phase. Pink, green
and yellow curves correspond to qualitatively different trajectories
obtained by our variational method. Attractive fixed-points are depicted
by black dots and the red dots represent unstable or saddle points.
An example of the dynamics for a finite-$s$, obtained by exact diagonalization
of the Liouvillian, is depicted as blue dashed lines and the steady-state
attained in the limit $t\to\infty$ is depicted as blue dots.

\subsubsection*{Phases\label{subsec:Phases-par}}

- Region $\text{I}_{\parallel}$ is characterized by a unique stable
steady-state located along the $z$-axis. The average magnetization
of the steady-state for finite-$s$ approaches the variational ansatz
value up to $1/s$ corrections (almost coinciding blue and black points
in Fig. \ref{fig:pd_pz}-$\text{I}_{\parallel}$). The variational
and finite-$s$ dynamics (green and blue-dashed lines respectively)
yield qualitatively similar results. In addition to the attractive
fixed point at the south pole (black dot), an unstable fixed point
is located at the north pole (red dot). Saddle points, not present
for the choice of parameters of Fig. \ref{fig:pd_pz}-$\text{I}_{\parallel}$,
may appear but do not change the dynamics qualitatively .

- In region $\t{II}_{\parallel}$ (Fig. \ref{fig:pd_pz}-$\t{II}_{\parallel}$),
we find two variational steady-states related by symmetry. For finite-$s$,
the degeneracy of the eigenvalues of the Liouvillian is lifted and
a unique steady-state emerges (blue dot) whose magnetization approaches
the average of the two variational ones. In the variational dynamics,
one of the two attractors is attained at large times depending on
the initial condition (green and pink lines in Fig. \ref{fig:pd_pz}-$\t{II}_{\parallel}$);
for finite-$s$ (blue dashed line) there are two separated time scales,
the initial dynamics approaches one of the variational fixed points
and is followed by a decay to the finite-s steady state. We analyze
the two time scale dynamics in Sec. \ref{subsec:Parallel-case}.

- Region $\t{III}_{\parallel}$ has three variational stable fixed
points (two related by symmetry and one with $\av{\bs S}=-s\bs e_{z}$).
Which fixed point is realized in the $t\to\infty$ limit depends on
the basin of attraction of the initial state. The finite-$s$ dynamics
also shows a separation of time-scales, similar to region $\t{II}_{\parallel}$,
before the finite-$s$ steady-state is attained.

\subsubsection*{Phase transitions\label{subsec:Phase-transitions-par}}

\begin{figure}
\includegraphics[width=1\columnwidth]{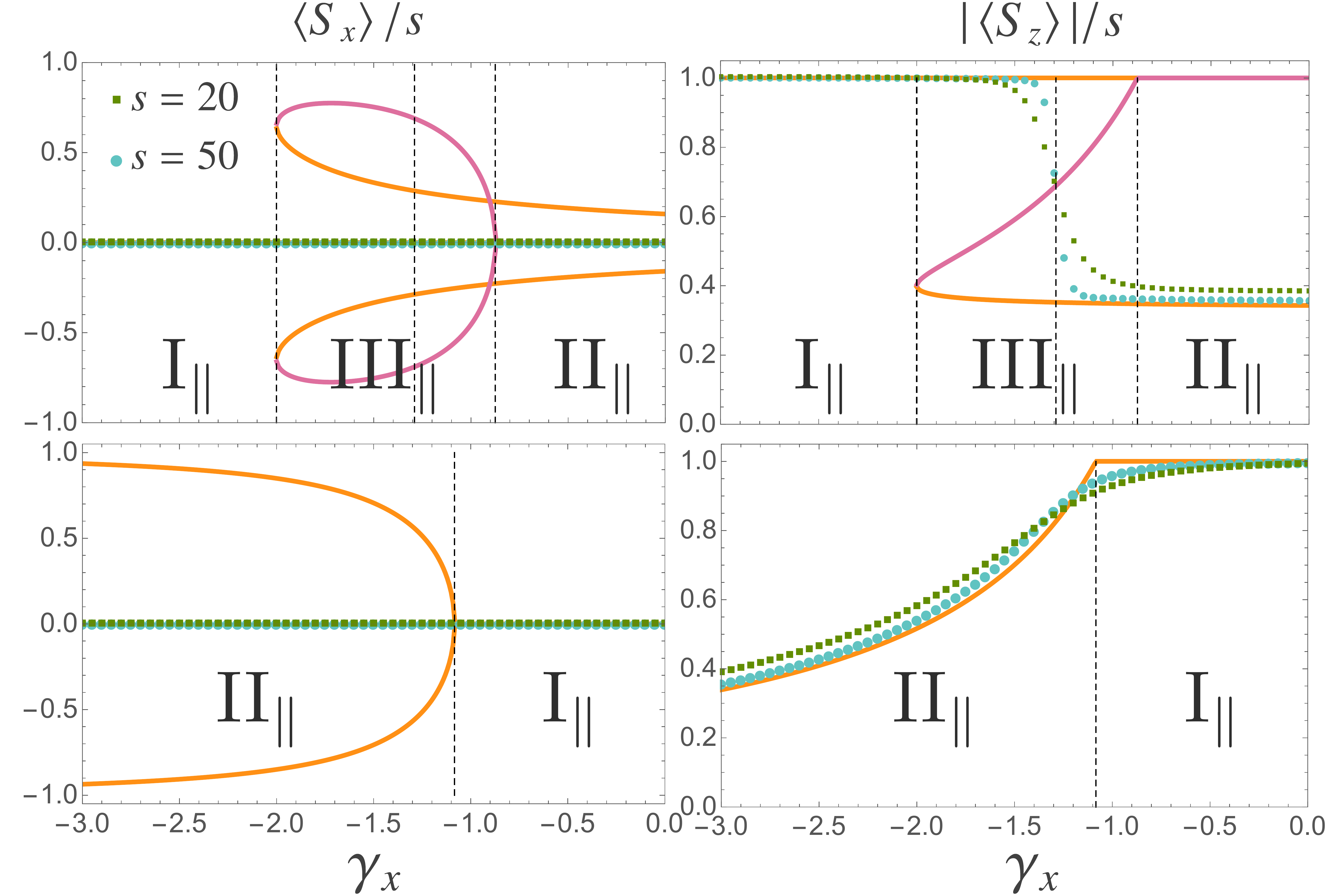}\caption{\label{fig:transition_pz}Upper panels: magnetization for the phase
transition $\protect\t I_{\parallel}\leftrightarrow\protect\t{III}_{\parallel}\leftrightarrow\protect\t{II}_{\parallel}$
for $h=1,\gamma_{y}=-3,p\Gamma=1$. Lower panels: magnetization for
phase transition $\protect\t I_{\parallel}\leftrightarrow\protect\t{II}_{\parallel}$
for $h=1,\gamma_{y}=2,p\Gamma=1$. The stable and unstable fixed points
are depicted as orange and pink lines respectively.}
\end{figure}

\begin{figure}
\includegraphics[width=1\columnwidth]{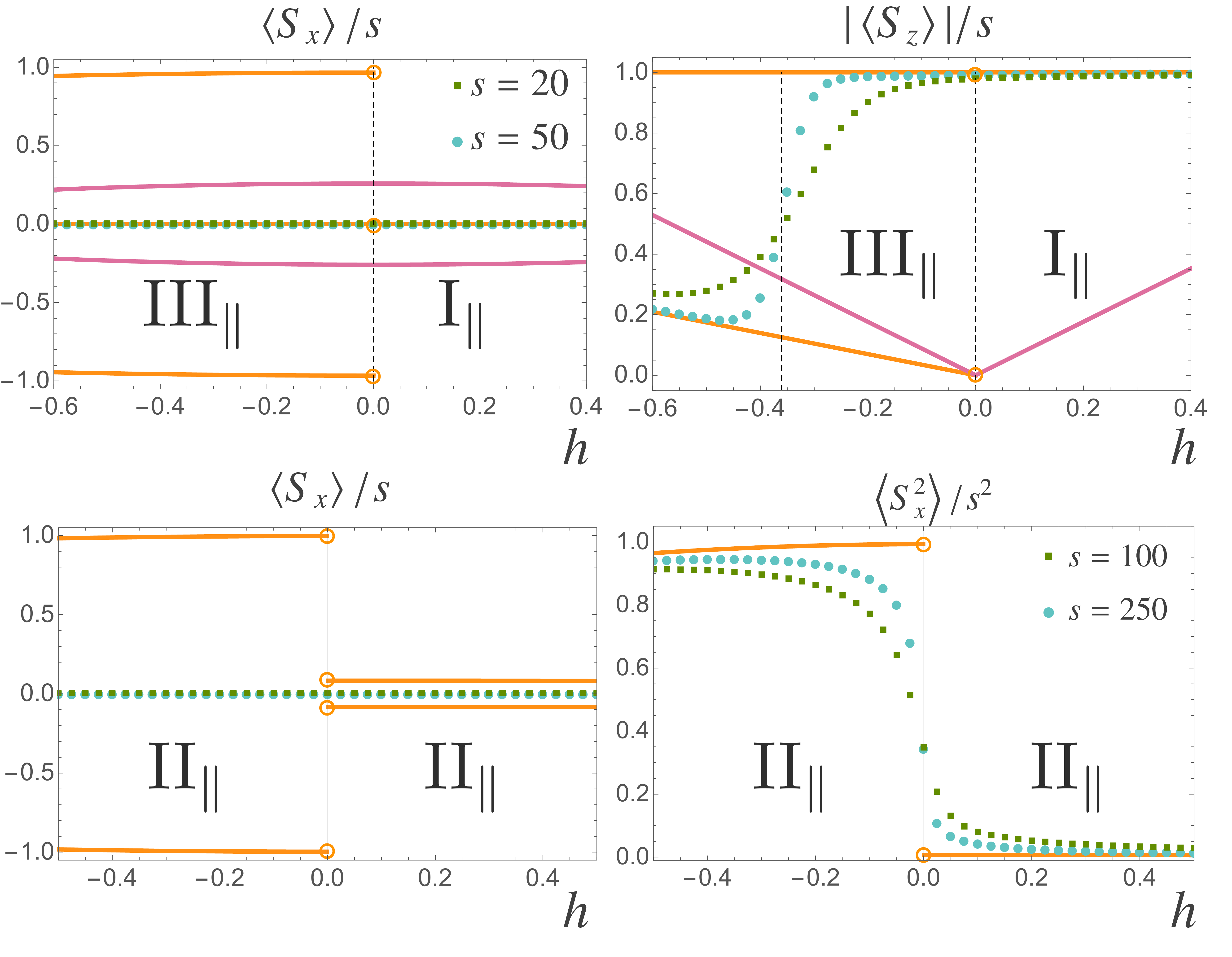}\caption{\label{fig:transition_pzh}Upper panels: magnetization at the phase
transition $\protect\t I_{\parallel}\leftrightarrow\protect\t{III}_{\parallel}$
with $\gamma_{y}=3,\gamma_{y}=1,p\Gamma=1$. Lower panels: magnetization
at the phase transition for $\protect\t{II}_{\parallel}\leftrightarrow\protect\t{II}_{\parallel}$
with $\gamma_{y}=3,\gamma_{y}=-3,p\Gamma=1$. The stable and unstable
infinite-$s$ results are depicted as orange and pink lines respectively.}
\end{figure}

We now turn to the description of the phase transitions. Figs. \ref{fig:transition_pz}
and \ref{fig:transition_pzh} show the magnetization in the $x$ (left
panels) and $z$ (right panels) directions as a function of $\gamma_{x}$
and $h$, for finite values of $s$ (blue and green dots) and for
the stable (orange) and unstable (pink) fixed-point of the variational
dynamics. Fig. \ref{fig:transition_pz} depicts the passage from phase
$\t I_{\parallel}$ to phase $\t{II}_{\parallel}$, with (upper panels)
and without (lower panels) the presence of the intermediate phase
$\t{III}_{\parallel}$. Fig. \ref{fig:transition_pzh} shows a cross
section of the phase diagram of Fig. \ref{fig:pd_pz}-(left panels)
obtained by varying $h$ along two vertical lines that cross the $\t I_{\parallel}\leftrightarrow\t{III}_{\parallel}$
transition (upper panels) and the $\t{II}_{\parallel}\leftrightarrow\t{II}_{\parallel}$
one (lower panels) that crosses the $\t 0_{\parallel}$ critical plane.


Analyzing Figs. \ref{fig:transition_pz} and \ref{fig:transition_pzh},
we list the properties of each transition.

- The transition $\t I_{\parallel}\leftrightarrow\t{II}_{\parallel}$
is of second order, with the unique steady-state of $\t I_{\parallel}$
giving place to two symmetry broken ones for $\t{II}_{\parallel}$
{[}see Fig. \ref{fig:transition_pz}-(lower panels)\textcolor{black}{{]}.}\textcolor{red}{{}
}A good order parameter for this transition is $\av{S_{z}}/s-1$,
which vanishes in phase $\t I_{\parallel}$ and is non-zero in phase
$\t{II}_{\parallel}$.

- At the $\t{II}_{\parallel}\leftrightarrow\t{III}_{\parallel}$ and
$\t{III}_{\parallel}\leftrightarrow\t I_{\parallel}$ transitions,
the quantity $\lim_{s\to\infty}\av{S_{z}}/s$ computed in the steady-state
is analytic as seen in Fig. \ref{fig:transition_pz}-(upper panels).
Analyticity was also observed for all other steady-state observables.
Therefore, these transitions only concern dynamic properties. 

- A discontinuous steady-state phase transition arises within $\t{III}_{\parallel}$.
For finite-$s$, quantum fluctuations select a steady-state with an
average magnetization that is either that of the stable fixed-point
of $\t I_{\parallel}$ or the average of the fixed-points of $\t{II}_{\parallel}$,
since these three fixed-points coexist in $\t{III}_{\parallel}$.
This scenario of a first order phase transition is similar to that
reported in Ref. \citep{Casteels2017}, the only difference being
that the phase equivalent to $\t{II}_{\parallel}$ has in Ref. \citep{Casteels2017}
a unique stable fixed-point.

- The transition $\t I_{\parallel}\leftrightarrow\t{III}_{\parallel}$
across the plane $h=0$ is of first order. However, since the symmetry
is not broken for finite-$s$, the steady-state magnetization is continuous,
see Fig. \ref{fig:transition_pzh}.

- The transition $\t{II}_{\parallel}\leftrightarrow\t{II}_{\parallel}$
across the $h=0$ plane is also of first order. The discontinuity
of $\av{S_{x}^{2}}/s^{2}$ is shown in Fig. \ref{fig:transition_pzh}.

\subsection{Perpendicular polarization\label{subsec:Perpendicular-polarization-III}}

\textcolor{black}{The case $\bs p\perp\bs h$ shown in Fig. \ref{fig:pd_py}-(left
panel), has three different phases: $\t 0_{\perp}$, $\t I'_{\perp}$
and $\t I_{\perp}$. The corresponding dynamics is plotted in Fig.
\ref{fig:pd_py}-(right panels) with the same color code of Fig. \ref{fig:pd_pz}.
In addition, the gray line in Fig. \ref{fig:pd_py}-$\t I'_{\perp}$
depict}s a separatrix curve dividing orbits where variational ansatz
has qualitatively different dynamics. Note that, both $\t 0_{\perp}$
and $\text{I}'_{\perp}$ support states that do not relax in the infinite-$s$
limit.

\subsubsection*{Phases\label{subsec:Phases-perp}}

- Region $\text{0}_{\perp}$ has no variational stable fixed-points.
However, the variational method finds a line of marginal fixed-point
solutions (brown line) where the eigenvalues of the stability matrix,
obtained by linearizing the equations of motion, have a zero real
part. This line connects two marginal steady-states that satisfy $\abs{\bs n}=1$,
depicted as red dots on the $z=0$ plane in Fig. \ref{fig:pd_py}-$\text{0}_{\perp}$\textcolor{black}{.}\textcolor{red}{{}
}The dynamics of any initial condition (green and pink lines) follows
closed orbits that surround the marginal line. Thus, the asymptotic
long-time state of the variational dynamics is recurrent and keeps
memory of the initial condition for all times. The existence of recurrent
classical solutions was previously identified in \citep{Kilin1978,Drummond1978,Drummond1980,Carmichael1999}
and recently studied in \citep{Hannukainen2017,Iemini2017}. For the
case $\gamma_{x}=\gamma_{y}=0$ and $p=-1$ an explicit solution of
the steady-state for finite-$s$ is known \citep{Kilin1978,Drummond1978,Drummond1980}.

For finite-$s$, a single unique steady-state (blue dot), with $\abs{\av{\bs S}}/s\neq1$,
is attained. This fixed-point corresponds to the unique place along
the line of marginal fixed-points where $\av{S_{x}}=0$, which is
consistent with the fact that the finite-$s$ steady-state cannot
break the microscopic symmetries.

The finite-$s$ picture emerging from our variational dynamics is
the following: finite size corrections destabilize the recurrent variational
evolution (valid for $s\to\infty$) and, after a timescale that increases
with $s^{-1}$ (see Sec.\ref{sec:Dynamics-for-finite-s}), the unique
steady-state is attained. Note that, if the initial state is arbitrarily
close to one of the marginal fixed points, the evolution to the finite-$s$
steady-state is along the lines of marginal fixed-points found by
the variational method. Therefore, including $1/s$ corrections to
the variational procedure is expected to lift the degeneracy of the
states along the line and yield a unique steady-state that coincided
with the finite-$s$ one.

- Region $\text{I'}_{\perp}$ is characterized by a stable fixed-point
solution coexisting with recurrent states. A separatrix line ( gray
line in Fig. \ref{fig:pd_py}-$\t I'_{\perp}$) separates the region
where an initial state attains asymptotically the stable fixed point
(e.g. green trajectory) from the region where an initial state yields
a recurrent evolution (e.g. pink trajectory). The finite-$s$ evolution
(blue dashed line) starting from an initial state in the recurrent
region, first follows the variational recurrent evolution and, subsequently,
decays towards the unique stable fixed-point.

- Region $\t I_{\perp}$ has a single stable steady-state and the
same qualitative properties as $\text{I}_{\parallel}$. This region
exists only for $h<h_{c}=p\Gamma/2$.

\subsubsection*{Phase transitions\label{subsec:Phase-transitions-perp}}

\begin{figure}
\includegraphics[width=1\columnwidth]{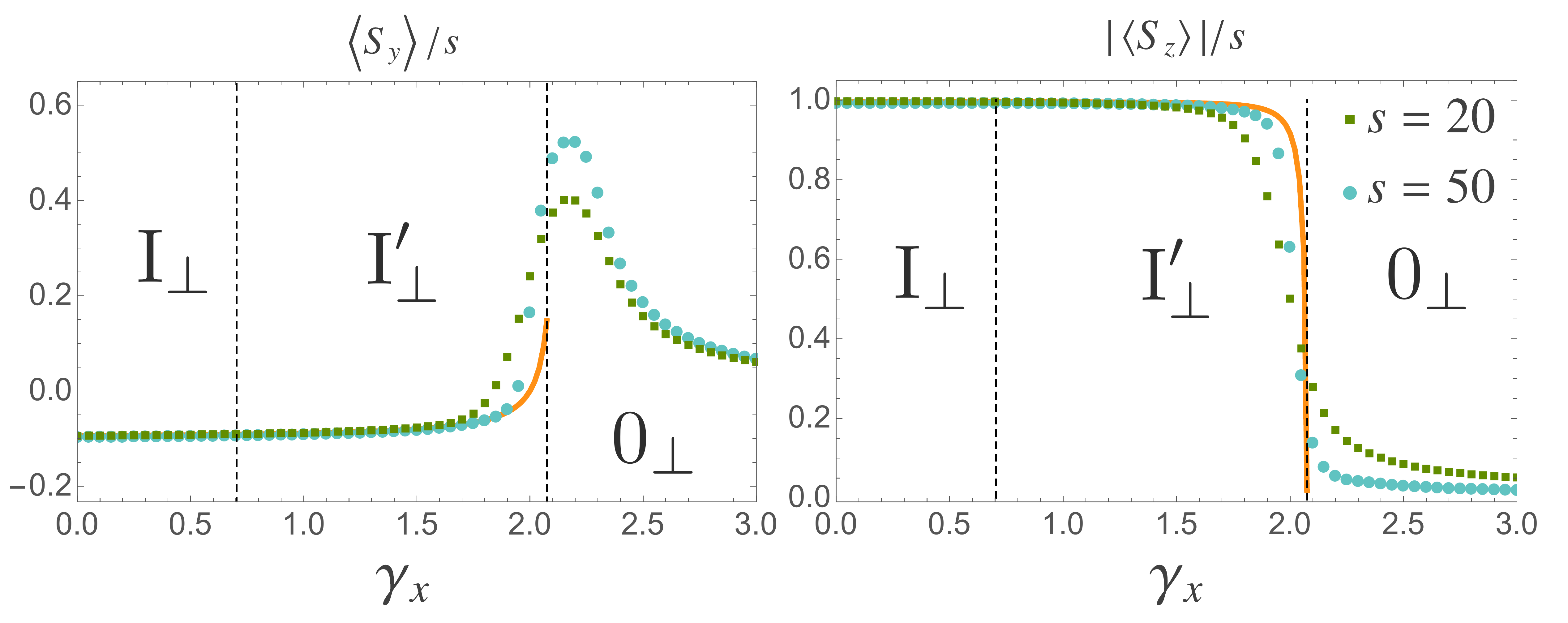}\caption{\label{fig:transition_py}Magnetization across the $\text{0}_{\perp}\leftrightarrow\protect\t I'_{\perp}\leftrightarrow\protect\t I_{\perp}$
transitions for $h=0.2,\gamma_{y}=-2,p\Gamma=1$. The stable infinite-$s$
steady-state is depicted as orange line.}
\end{figure}

The phase transitions in the perpendicular case can be of two kinds
$\text{0}_{\perp}\leftrightarrow\t I'_{\perp}$ and $\t I'_{\perp}\leftrightarrow\t I_{\perp}$.
Fig. \ref{fig:transition_py} shows the magnetization in the $y$
(left panel) and $z$ (right panel) directions as a function of $\gamma_{x}$
for two values of $s$ (blue and green dots) and for the stable fixed-point
obtained by the variational ansatz (orange curve). When $h<1/2$,
there are two points within a fixed $h$ plane for which the passage
from $\text{0}_{\perp}$ to $\t I_{\perp}$ can be done directly,
without passing by $\t I'_{\perp}$. As the steady-state properties
of phases $\t I'_{\perp}$ and $\t I_{\perp}$ are similar, crossing
the transition along these special points will not affect qualitatively
the scenario presented in Fig. \ref{fig:transition_py}.

- The $\text{0}_{\perp}\leftrightarrow\t I'_{\perp}$ transition is
of first order, with a discontinuous magnetization shown in Fig. \ref{fig:transition_py}.
However, as there is no stable fixed-point within phase $\text{0}_{\perp}$,
this transition seems to escape the Landau paradigm \citep{Hannukainen2017}.

- The $\t I'_{\perp}\leftrightarrow\t I_{\perp}$ transition regards
only the spectral properties of the Liouvillian and is discussed below.
The steady-state magnetization, depicted in Fig. \ref{fig:transition_py}
for finite $s$, is continuous across the transition for $s\to\infty$. 

\section{Steady-State, Spectral and Dynamic Signatures of Non-Equilibrium
Phases\label{sec:Linearized-Liouvillian-operator}}

In this section, we analyze the spectral and steady-state properties
of the phases described in Sec. \ref{sec:Phase-Diagram}. For these
quantities, large-$s$ predictions require to go beyond the variational
analysis. We achieve this using a Holstein-Primakoff transformation,
mapping the spin into a bosonic degree of freedom, which allow a subsequent
$1/s$ expansion of the Liouvillian. At leading order, the bosonic
Liouvillian is quadratic and thus exactly solvable. Details of the
exact solution are given in Appendix \ref{sec:The Linearized Lindblad Operator}.
Analytic predictions obtain in this way are then compared with exact
diagonalization results.

The main findings of this section are summarized in columns 4 and
5 of Table \ref{tab:Classification-of-steady-state} and discussed
in Sec. \ref{subsec:Classification-of-steady-state}.

\subsection{Holstein-Primakoff transformed Liouvillian\label{subsec:Holstein-Primakoff-transformed-L}}

The Holstein-Primakoff (H-P) transformation maps a spin$-s$ into
a bosonic degree of freedom. A generalized version of this transformation,
which conserves the spin commutation relations, can be obtained by
the usual mapping
\begin{align}
S_{z} & =-s+a^{\dagger}a\\
S_{+} & =a\sqrt{2s-a^{\dagger}a}\\
S_{-} & =\sqrt{2s-a^{\dagger}a}\,a^{\dagger}
\end{align}
followed by a shift in the bosonic operators $a\to a+\sqrt{2s}\frac{\alpha}{\sqrt{1+\bar{\alpha}\alpha}}$
, with $\alpha\in\mathbb{C}$. This generalized H-P mapping allows
a systematic $1/s$ development around a spin-coherent state, $\ket{\alpha}_{c}=e^{\alpha S_{+}}\ket{s,-s}$,
parametrized by $\alpha$, with average magnetization

\begin{align}
\av{\bs S} & =s\left\{ \frac{\alpha+\text{\ensuremath{\bar{\alpha}}}}{1+\bar{\alpha}\alpha},i\frac{\alpha-\bar{\alpha}}{1+\bar{\alpha}\alpha},1-\frac{2}{1+\bar{\alpha}\alpha}\right\} +O\left(\sqrt{s}\right).\nonumber \\
\end{align}
Inserting the expansion of the spin operators in the Liouvillian and
developing in powers of $s$, up to order $s^{0}$, we obtain a quadratic
Liouvillian in the bosonic operators, where $H$ can generically be
casted in the form
\begin{align}
H & =A^{\dagger}.\bs H.A+A^{\dagger}.\zeta+\zeta^{\dagger}.A+O\left(s^{-1/2}\right)
\end{align}
with $A=\left\{ a,a^{\dagger}\right\} ^{T}$, the single-particle
Hamiltonian $\bs H$ is a $2\times2$ matrix and $\zeta$ a two-component
complex vector. In the same way the jump operator $W_{i}$ can be
written as 
\begin{align}
W_{i} & =w_{i}^{\dagger}.A+c_{i}
\end{align}
with $w_{i}$ a two-component complex vector and $c_{i}$ a complex
constant. The quantities $\bs H$ and $w_{i}$ are of order $s^{0}$
and $\zeta$ and $c_{i}$ are of order $s^{1/2}$. A suitable choice
of the shift, $\alpha$, can be used to set to zero the terms proportional
to $\zeta$ or $c_{0}$ in the linearized Liouvillian, obtaining an
operator with only quadratic terms. The values of $\alpha$ that have
this property are those that fulfill fixed-point conditions of the
variational and semi-classical dynamics given in Appendix \ref{sec: Semiclassical dynamics}.
This step, is thus, equivalent to choose as linearization points the
fixed points of the infinite-$s$ equation of motion with $\abs{\bs n}=1$.

Properties of quadratic bosonic Liouvillians were studied in Ref.
\citep{Prosen2010b}. We derive some of these results in the Appendix
\ref{sec:The Linearized Lindblad Operator} using an approached similar
to that developed in Ref. \citep{Ribeiro2014e} for quadratic fermionic
Liouvillians. Using this method, we compute the single particle correlation
matrix, $\bs{\chi}=\av{A.A^{\dagger}}$, which encodes the properties
of the steady-state, the spectral gap, and derive the simple structure
of the low energy spectrum.

\subsection{Steady-state\label{subsec:Steady-state}}

\begin{figure}
\includegraphics[width=1\columnwidth]{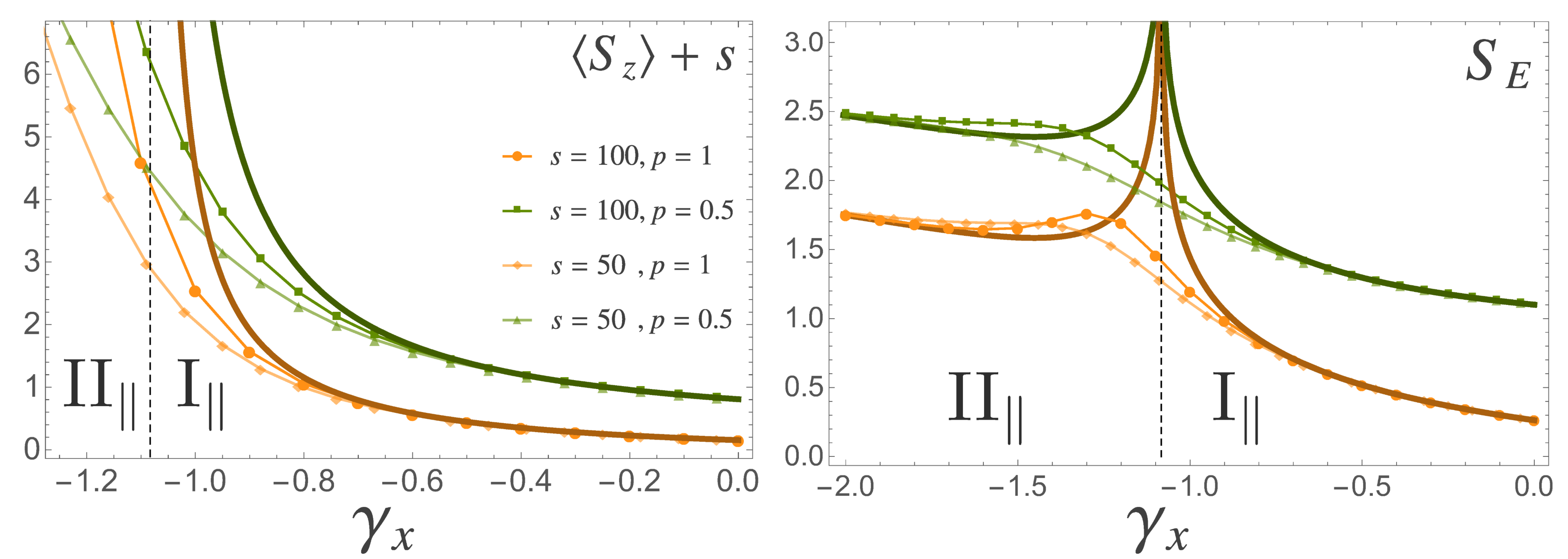}

\caption{\label{fig:ss_pz}$1/s$ corrections to $\protect\av{S_{z}}-s$ (left
panel) and $S_{E}$ (right panel) with $h=1,\gamma_{y}=2,\Gamma=1/p$.}
\end{figure}

In this section, we study steady-state properties starting with the
parallel polarization case ($\bs p\parallel\bs h$).

For phase $\text{I}_{\parallel}$, there is only one stable solution,
$\alpha_{1}$, of the variational equations, thus to leading order
in $s$, $\rho_{0}\simeq\ket{\alpha_{1}}_{c}\bra{\alpha_{1}}_{c}$.
Analytic predictions for the steady-state observables to next-to-leading
order can be obtained using density matrix $\rho_{0}=\chi_{1}$, where
$\chi_{1}$ is the density matrix obtained by linearizing the Liouvillian
around $\alpha_{1}$.

For phase $\text{II}_{\parallel}$, at leading order in $s$, $\mathcal{L}$
has two eigenstates with eigenvalues exponentially close to zero that
are well approximated by $\rho_{0}=\frac{1}{Z_{0}}\left(\ket{\alpha_{1}}_{c}\bra{\alpha_{1}}_{c}+\ket{\alpha_{2}}_{c}\bra{\alpha_{2}}_{c}\right)$,  
with $Z_{0}=\tr\left(\ket{\alpha_{1}}_{c}\bra{\alpha_{1}}_{c}+\ket{\alpha_{2}}_{c}\bra{\alpha_{2}}_{c}\right)$,  
and $\rho_{1}=\ket{\alpha_{1}}_{c}\bra{\alpha_{1}}_{c}-\ket{\alpha_{2}}_{c}\bra{\alpha_{2}}_{c}$,  
from which only $\rho_{0}$ is a physical density matrix. At next
to leading order in $s$, the density matrix is given by $\rho_{0}=\frac{1}{2}\left(\chi_{1}+\chi_{2}\right)$,
where $\chi_{1,2}$ are the finite entropy density matrices obtained
by linearizing the Liouvillian around $\alpha_{1,2}$, respectively.
Since the overlap $\braket{\alpha_{1}}{\alpha_{2}}_{c}$ is exponentially
small in $s$, $\chi_{1}$ and $\chi_{2}$ are exponentially non-overlapping,
i.e. $\ln\tr\left(\chi_{1}\chi_{2}\right)\propto-s$. As a consequence,
mean values of operators can be approximated by $\tr\left(\rho_{0}O\right)\simeq\frac{1}{2}\left[\tr\left(\chi_{1}O\right)+\tr\left(\chi_{2}O\right)\right]$.
The entropy of $\rho_{0}$ is also well approximated by $S_{\text{E}}\simeq\ln2-\frac{1}{2}\tr\left(\chi_{1}\ln\chi_{1}\right)-\frac{1}{2}\tr\left(\chi_{2}\ln\chi_{2}\right)=\ln2-\tr\left(\chi_{1}\ln\chi_{1}\right)$,
since by symmetry the entropy of $\chi_{1}$ and $\chi_{2}$ are equal.

Fig. \ref{fig:ss_pz} shows, the $1/s$ corrections to the magnetization
$\av{S_{z}}-s$ and the von Neumann entropy, $S_{\text{E}}=-\tr\left(\rho\ln\rho\right)$,
of the steady-state as a function of $\gamma_{x}$, in phases $\text{I}_{\parallel}$
and $\text{II}_{\parallel}$ and across the $\text{I}_{\parallel}\leftrightarrow\text{II}_{\parallel}$
transition. Since in phase $\t I_{\parallel}$, the magnetization
satisfies $\av{S_{z}}=-s+\delta S_{z}+O\left(s^{-1}\right)$, the
values of $\delta S_{z}=s+\av{S_{z}}$ for finite-$s$ converge to
the analytic predictions obtained using the linearized Liouvillian
around the stable steady-state. For the entropy, Fig. \ref{fig:ss_pz}
shows that the numerical results tend to the analytic predictions
as $s\to\infty$. The convergence is much slower around the phase
transition point.

At the phase transition, the perturbative expansion is no longer valid
and the above estimate breaks down. When the linearized steady-state
is a good approximation of the finite-$s$ one, the von Neumann entropy
in the $s=\infty$ limit approaches a constant value. The proximity
with the critical point where the linearized procedure breaks down,
explains the slow convergence with $s$.

For the perpendicular polarization case ($\bs p\perp\bs h$) and in
the regions where a stable steady-state is present ($\t I_{\perp}$and
$\t I'_{\perp})$, the properties of the steady-state are similar
to those of region $\text{I}_{\parallel}$. On the other hand, the
recurrent region $0_{\perp}$ has no stable fixed-point to approximate
the finite-$s$ steady-state. In this case, as presented below, the
entanglement entropy of the finite-$s$ steady state grows as $\ln(s)$.
It is tempting to interpret this logarithmic growth as an extension
of the argument above for phase $\text{II}_{\parallel}$, where $O\left(s\right)$
degenerate steady-states contribute equally to $S_{\text{E}}$.

\subsection{Spectrum and characteristic time-scales\label{subsec:Spectrum and dynamics}}

We now focus on spectrum of the Liouvillian linearized around each
steady-state. For the case of a single bosonic mode obtained by $1/s$
expansion of the H-P transformation, the eigenvalues $\Lambda_{n,m}$
of the Liouvillian are given by $\Lambda_{n,m}=i\left(n\lambda-m\bar{\lambda}\right)$,
with $m,n\in\mathbb{N}_{0}^{+}$ , where $\lambda$ is a complex number
that can be obtained from $\bs H$ and $w_{i}$ (see Appendix \ref{sec:The Linearized Lindblad Operator}
). Each eigenvalue corresponds to a decaying mode of the dynamics
towards the steady-state with a characteristic time scale $\tau=-\left(\re\Lambda\right)^{-1}$.

\subsubsection*{Parallel case\label{subsec:Parallel-case}}

\begin{figure}
\includegraphics[width=1\columnwidth]{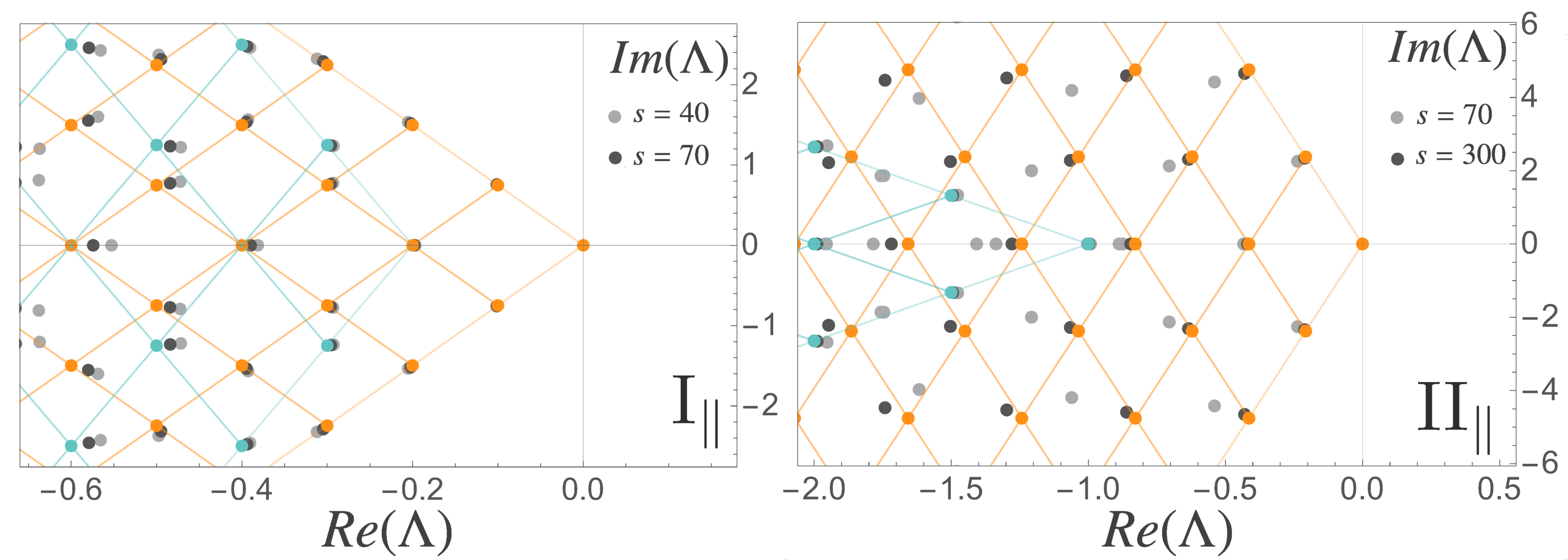}

\caption{\label{fig:sp_pz} Numerical eigenspectrum of $\mathcal{L}$ for regions
$\text{I}_{\parallel}$ (left) and $\text{II}_{\parallel}$ (right)
and analytical predictions obtained from the linearized Liouvillian
around the stable fixed point (light orange) and around the unstable
one (light blue). Parameters from Fig.\ref{fig:pd_pz}.}
\end{figure}

Fig. \ref{fig:sp_pz} depicts the spectrum of the Liouvillian $\mathcal{L}$
in $\t I_{\parallel}$ and $\t{II}_{\parallel}$ . The gray level
coded dots correspond to spectrum of the full Liouvillian for increasing
values of $s$. The orange (blue) dots correspond to the spectrum
of the linearized Liouvillian around the stable (unstable) fixed points,
the orange (blue) lines were drawn to highlight the simple periodic
structure of the spectrum.

For the case of Fig. \ref{fig:sp_pz}-$\t I_{\parallel}$, the spectrum
is generated by $\lambda=\lambda_{\t s}=\sqrt{\left(h+\gamma_{x}\right)\left(h+\gamma_{y}\right)}-\frac{1}{2}i\Gamma p$
(see derivation in Sec. \ref{subsec:Explicit-example:-Region}). Note
that the agreement between the finite-$s$ spectrum and the linearized
one is faster for small values of $\abs{\re\left(\Lambda_{n,m}\right)}$.
For larger values, we can still observe a convergence to the linearized
prediction with increasing $s$. The decay towards the unique steady-state,
after the fast decaying modes vanish, is ruled by the two slowest
decaying modes depicted in Fig. \ref{fig:sp_pz}-$\text{I}_{\parallel}$
with a characteristic time-scale $\tau_{0}=\abs{\im\left(\lambda_{s}\right)}^{-1}$.

In case of Fig. \ref{fig:sp_pz}-$\t{II}_{\parallel}$ there are two
stable fixed points related by symmetry. Linearizing the Liouvillian
around each of these fixed points yields a spectrum that is doubly
degenerate. A quasi-degeneracy is also observed in the finite-$s$
spectrum obtained by exact diagonalization with a convergence to the
linearized prediction with increasing $s$.

In region $\t{II}_{\parallel}$, the dynamics for finite-$s$ is thus
characterized by two different time scales. The first timescale, of
order $s^{0}$, is given by $\tau_{0}=\abs{\im\left(\lambda_{s}\right)}^{-1}$,
with $\lambda_{\t s}$ obtained by linearizing the Liouvillian around
one of the two symmetry-related stable steady-states. \textcolor{black}{The
choice of the particular steady-state depends on which basin of attraction
the initial conditions belong to. Within this timescale, the evolution
of a finite-$s$ system ten}ds to the infinite-$s$ evolution as the
value of $s$ increases. For times $t>\tau_{0}$, the dynamics resolves
the degeneracy between the steady-state $\rho_{0}$ and the first
excited state $\rho_{1}$ of $\mathcal{L}$ defined in Sec. \ref{subsec:Steady-state}
and the decay is dominated by the inverse of the first non-zero eigenvalue
$\Lambda_{1}$ of $\mathcal{L}$, $\tau_{1}=-(\re\Lambda_{\t 1})^{-1}$.
As $\re\left(\Lambda_{1}\right)$ is exponentially small in $s$,
these two timescales become increasingly separated for large $s$
and can be well identified in the dynamics (see Appendix \ref{sec:Dynamics-for-finite-s}
for more details).

The spectrum of region $\t{III}_{\parallel}$ is thrice degenerate
in the infinite-$s$ limit and we also observe convergence as $s$
increases (plot not shown). The dynamics in the region is similar
to phase $\t{II}_{\parallel}$ with the exception that now there are
three relevant time-scales. The first, $\tau_{0}=\abs{\im\left(\lambda_{s}\right)}^{-1}$,
determines the convergence to the basin dependent steady-state. One
of the two other timescales ($\tau_{1}$ or $\tau_{2}$) corresponds,
as in phase $\t{II}_{\parallel}$, to the decay from one of the symmetry
related states to the symmetric mixed-state. The other, to the decay
between the mixed-symmetric state and a state with $\av{\bs S}\simeq-s\bs e_{z}$
(as the steady-state of $\t I_{\parallel}$). Which eigenvalue, $\Lambda_{1}$
or $\Lambda_{2}$, corresponds to each of these processes depends
on what side of the first order transition the system is in.

Interestingly, there is a set of low-lying eigenvalues (blue dots)
obtained by exact diagonalization that do not converge to the spectrum
of the bosonic Liouvillian linearized around the stable fixed points.
Instead, these second set of eigenvalues can be obtained by linearizing
the Liouvillian around the unstable fixed-points. This spectrum has
a similar structure (blue lines) to that of the stable fixed point
but the element with the smallest real part within this set of eigenvalues
has a finite negative value, i.e. it is not a steady-state. For the
case $\t I_{\parallel}$, we obtain $\lambda=\lambda_{\t{uns}}=\sqrt{\left(h-\gamma_{x}\right)\left(h-\gamma_{y}\right)}+\frac{i\Gamma p}{2}$
(see derivation in Sec. \ref{subsec:Explicit-example:-Region}) and
the cone-like structure is displaced from the real axis by $-\Gamma p$.
A convergence to this second set of analytical predictions is also
observed in cases $\t I_{\parallel}$ and $\t{II}_{\parallel}$.

Therefore, the lower part of the spectrum of the full Liouvillian,
that rules the long-time dynamics, is an overlap of the spectra of
linearized Liouvillians around both stable and unstable fixed points.
Thus, in addition to the characteristic timescales determined by the
stable fixed-points, the long-time dynamics also carries information
about the unstable fixed points\textcolor{green}{.}

We now focus on the spectrum at the phases transitions of the parallel
case. As noticed before, there are three kinds of steady-state phase
transitions in the system: two first order, one with coexisting stable
fixed points ($\t I_{\parallel}\leftrightarrow\t{III}_{\parallel}\leftrightarrow\t{II}_{\parallel}$)
and one with no coexistence ($\t{II}_{\parallel}\leftrightarrow\t{II}_{\parallel}$),
and a second order phase transition ($\t I_{\parallel}\leftrightarrow\t{II}_{\parallel}$).
The $\t I_{\parallel}\leftrightarrow\t{III}_{\parallel}\leftrightarrow\t{II}_{\parallel}$
transition is hard to locate numerically and an analytical treatment
of the spectral properties beyond the heuristic picture given above
requires a non-perturbative treatment that is out of the scope of
this work. The transition $\t{II}_{\parallel}\leftrightarrow\t{II}_{\parallel}$
is realized passing by the $0_{\parallel}$ critical plane in Fig.
\ref{fig:pd_pz}-(left panel); the spectral and the steady-state properties
of this phase are similar to those of phase $0_{\perp}$ and will
be analyzed in the next section.

\begin{figure}
\includegraphics[width=1\columnwidth]{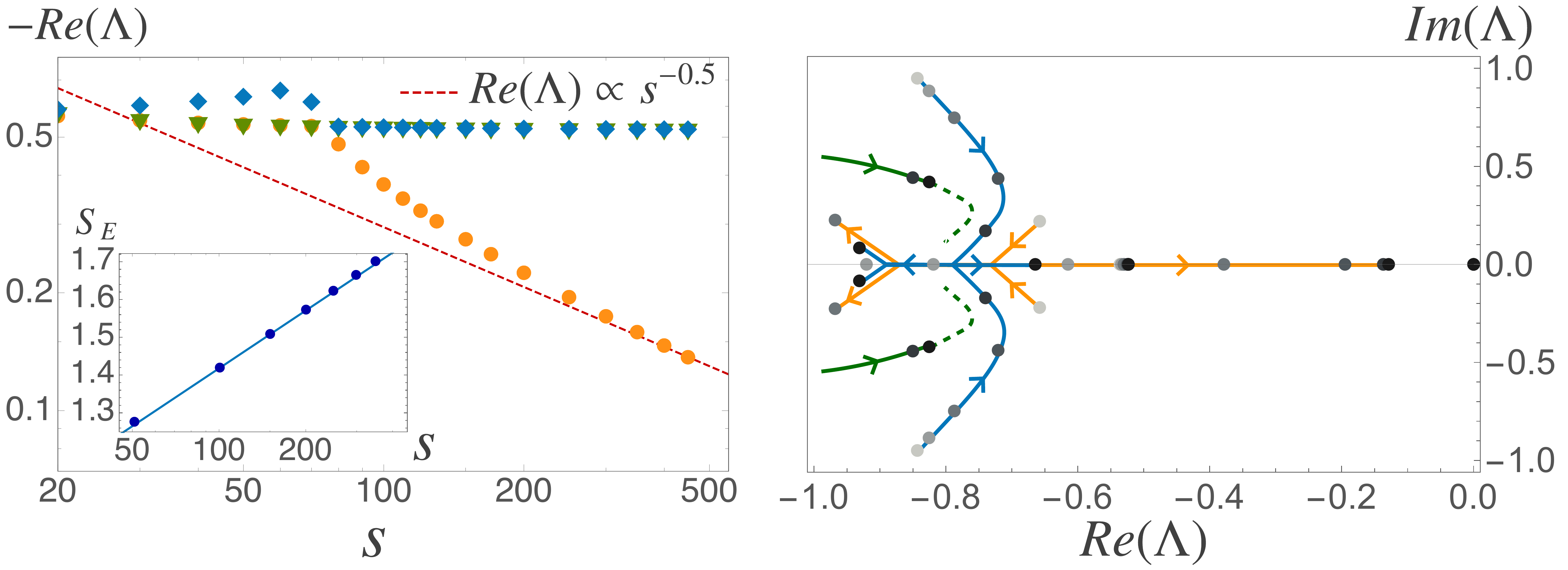}

\caption{\label{fig:lam7} Left panel: Real part of the smallest eigenvalues
at the transition $\protect\t I_{\parallel}\leftrightarrow\protect\t{II}_{\parallel}$
as a function of $s$, parameters of Fig.\ref{fig:pd_pz}. Inset:
von Neumann entropy of the steady-state as function of $s$. Right:
Spectrum of $\mathcal{L}_{\text{}}$ at the $\protect\t I_{\parallel}\leftrightarrow\protect\t{II}_{\parallel}$
transition, gray dots represent numeric values from $s=60$ up to
$s=500$ and the lines are trajectories as function of $s$.}
\end{figure}

The spectrum at the $\t I_{\parallel}\leftrightarrow\t{II}_{\parallel}$
critical point is depicted in Fig. \ref{fig:lam7}. As $s$ increases,
a larger number of eigenvalues approaches zero following a process
sketched in Fig. \ref{fig:lam7}-(right panel): for increasing s (see
arrows), two complex conjugate eigenvalues meet and become real; after
that one eigenvalue approaches zero. The behavior of the first\textcolor{green}{{}
}eigenvalue of the Liouvillian that converges to $0$ with $s$ is
given in Fig. \ref{fig:lam7}-(right panel), showing $\Lambda_{1}\propto s^{-\nu}$
with $\nu\simeq0.5$. The entropy of the finite-$s$ steady-state
is given in the inset of Fig. \ref{fig:lam7}-(left panel). The scaling
seems to be logarithmic in $s$, i.e. $S_{\t E}\propto\ln\left(s\right)$.
Away from the phase transition points, all steady-states have a finite
entropy in the infinite $s$ limit.

\subsubsection*{Perpendicular case\label{subsec:Perpendicular-case}}

The spectrum and dynamics of the magnetization and the entropy in
phase $\t I_{\perp}$, is similar to that of phase $\t I{}_{\parallel}$
in the previous section. Therefore, we refer the reader to the discussion
of phase $\t I{}_{\parallel}$ (above) for the physical understanding
of that phase.

\begin{figure}
\includegraphics[width=1\columnwidth]{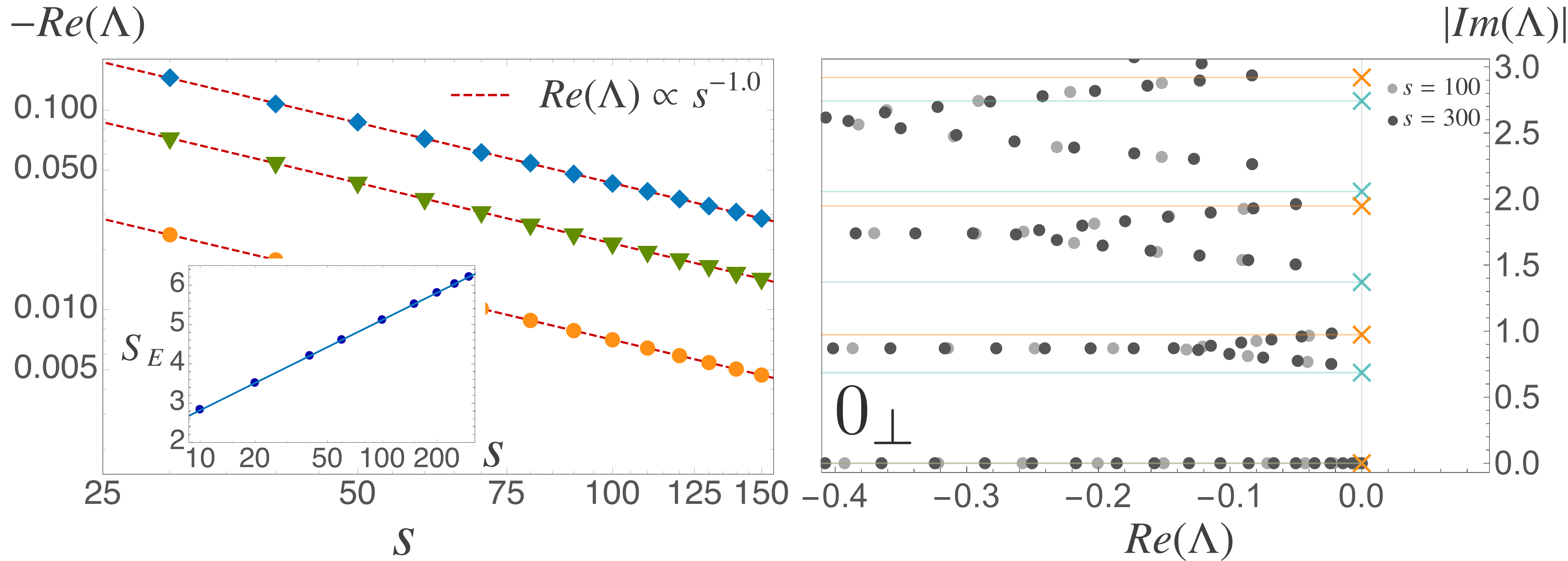}

\caption{\label{fig:sp_py_0}Left panel: Real part of the smallest eigenvalues
as a function of $s$ in region $\protect\t 0_{\perp}$, parameters
of Fig.\ref{fig:pd_pz}; Inset: von Neumann entropy as function of
$s$. Right panel: Spectrum of $\mathcal{L}$ for region $0_{\perp}$
, light (dark) gray dots represent numeric values for $s=100$ ($s=300$),
the orange and blue dots are obtained by linearizing the dynamics
around the two marginal infinite-$s$ fixed-points located on the
sphere.}
\end{figure}

Phases $\t 0_{\perp}$ and $\t I'_{\perp}$ allow for recurrent states
in the infinite-$s$ limit with a time-independent amplitude and frequency
which depend on the initial condition. In the $s\to\infty$ limit,
this corresponds to a spectrum of $\mathcal{\ensuremath{L}}$ with
an accumulation of points on the imaginary axis. This property, recently
studied in Ref. \citep{Iemini2017}, is shown in Fig. \ref{fig:sp_py_0}
for the case of a point in region $\t 0_{\perp}$. For finite-$s$
(see Fig. \ref{fig:sp_py_0}-right panel), we observe that some eigenvalues
indeed approach the imaginary axis, and, when sufficiently close to
the imaginary axis, fall along lines predicted for the marginal fixed
points of the linearized Liouvillian. In this phase, the Liouvillian
gap $\Lambda_{1}$ and the real part of the first few Liouvillian
eigenvalues ($\Lambda_{2}$, $\Lambda_{3}$,...) vanish as $s^{-1}$
(see Fig. \ref{fig:sp_py_0}-left panel).

This implies that in phase $\t 0_{\perp}$, the approach to the unique
finite-$s$ steady state is done with a rate of the order of $s^{-1}$.
The entropy of the finite-$s$ steady-state increases logarithmically
with $s$ (inset of Fig. \ref{fig:sp_py_0} - blue dots). In the same
inset we compare the entropy of state chosen by our variational procedure
(blue line) in Sec. \ref{subsec:Phases-perp} and find a remarkable
agreement (no fitting performed).

The spectrum in phase $\t I'_{\perp}$ (not shown) is a direct overlap
of the spectra of $\t 0_{\perp}$ and $\t I{}_{\perp}$ . Properties
of the finite-$s$ steady-state are always well approximated by a
quadratic Liouvillian, linearized around the stable fixed point of
$\t I'_{\perp}$.

\section{Classification of steady-state phases \label{subsec:Classification-of-steady-state}}

\begin{table*}
\begin{tabular}{|cl|c|c|c|c|}
\hline 
\multicolumn{2}{|c|}{Description} & Abbreviation & Region & Spectral Gap & S.S. Entropy\tabularnewline
\hline 
\hline 
\multicolumn{2}{|l|}{$\ $Non-Critical} &  &  &  & \tabularnewline
\hline 
$\ \ \ \ $ & Non-Degenerate & nCnD & $\t I_{\parallel},\t I_{\perp}$ & $\Delta\sim s^{0}$ & $S_{E}\sim s^{0}$\tabularnewline
\hline 
 & Degenerate - Symmetric & nCDS & $\t{II}_{\parallel}$ & $\ln\Delta\sim-s$ & $S_{E}\sim s^{0}$\tabularnewline
\hline 
 & Degenerate - Non-Symmetric & nCDnS & $\t{III}_{\parallel}$ & $\ln\Delta\sim-s$ & $S_{E}\sim s^{0}$\tabularnewline
\hline 
\multicolumn{2}{|l|}{$\ $Critical} &  &  &  & \tabularnewline
\hline 
 & Non-Recurrent & CnR & $\t I_{\parallel}\leftrightarrow\t{II}_{\parallel}$ & $\Delta\sim s^{-\frac{1}{2}}$ & $S_{E}\sim\ln s$\tabularnewline
\hline 
 & Coexistence & CC & $\t{I'}_{\perp}$,$\t I'_{\parallel}$ & $\Delta\sim s^{-1}$ & $S_{E}\sim s^{0}$\tabularnewline
\hline 
 & Recurrent & CR & $\t 0_{\perp}$,$\t 0_{\parallel}$ & $\Delta\sim s^{-1}$ & $S_{E}\sim\ln s$\tabularnewline
\hline 
\end{tabular}

\caption{\label{tab:Classification-of-steady-state}Classification of steady-state
phases.}
\end{table*}

We can now establish a complete classification of the different phases
of the model. A summary of the following discussion and acronyms table
is presented in Tab. \ref{tab:Classification-of-steady-state} and
should be understood as the main result in our paper.

We start by classifying the different systems in two major classes:
non-critical system (NCS), where the number of zero eigenvalues of
the Liouvillian operator is finite for $s\to\infty$; and critical
systems (CS) that have a spectrum where an infinite number of eigenvalues
approaches the imaginary axis as $s\to\infty$.

NCS correspond to the phases $\t I_{\parallel}$, $\t{II}_{\parallel}$,
$\t{III}_{\parallel}$ and $\t I_{\perp}$. For these systems, the
spectrum is well approximated by a linearized bosonic Liouvillian
obtained after a Holstein-Primakoff transformation around the (stable
and unstable) fixed points of the infinite-$s$ dynamics. Each stabl\textcolor{black}{e
point in the dynamics, $\alpha_{i=1,2,3}$, co}rresponds to a zero
eigenvalue on the Liouvillian in the $s\to\infty$ limit with an eigenvector
that is well approximated by the density matrix $\rho\simeq\ket{\alpha_{i}}_{c}\bra{\alpha_{i}}_{c}$,
with $\ket{\alpha_{i}}_{c}$ a spin coherent state. In NCS phases
with more than one infinite-$s$ steady-state, the dynamics follows
the two time-scale paradigm observed in phases $\t{II}_{\parallel}$,
$\t{III}_{\parallel}$. This corresponds to a first decay towards
the infinite-$s$ state in the basin of attraction of the initial
point, with a time scale of order $s^{0}$, and a second decay to
the finite-$s$ steady-state, with a time scale that diverges exponentially
as $s$ increases. Observables, such as the steady-state magnetization
and entropy, can be obtained, at every order in $s$, by systematically
computing $1/s$ corrections to the leading order linearized Liouvillian.
In particular, the von Neumann entropy is finite in the infinite-$s$
limit.

NCS systems can be divided into three sub-classes: non-degenerate
(nCnD), with unique single steady-state ($\t I_{\parallel},\t I_{\perp}$);
degenerate-symmetric (nCDS) and degenerate-nonsymmetric (nCDnS) where
more than one steady state exist ($\t{II}_{\parallel}$ and $\t{III}_{\perp}$
respectively).

- For nCDS phases, a pair of symmetry broken steady-states becomes
exponentially degenerate, $\Delta\sim\exp(-s)$, in the infinite-$s$
limit. Because the states break the symmetry of the underlying Liouvillian
in the infinite-$s$ limit, the finite-$s$ steady-state is well approximated
by a symmetric combination of the two infinite-$s$ states and we
say that the transition $\t I_{\parallel}\leftrightarrow\t{II}_{\parallel}$
is of second order.

- nCDnS phases can encompass multiple pairs of symmetry broken steady-states
and symmetric states. All steady-states are exponentially degenerate
in the infinite-$s$ limit, however in order to compute which of the
steady-states is realized for finite $s$, a non-perturbative calculation
in $s$ is needed that goes beyond the scope of the current work.

CS are represented in this work by regions $0_{\perp}$, $\t I'_{\perp}$
and by the phase transition planes, including: $\t 0_{\parallel}$,
$\t I'_{\parallel}$ and the transitions lines $\t I_{\parallel}\leftrightarrow\t{II}_{\parallel}$.
These can be divided into three sub-classes: recurrent (CR) with all
the initial states displaying recurrent behavior ($\t 0_{\perp}$,$\t 0_{\parallel}$),
coexistence (CC) whose properties depend on the initial state ($\t{I'}_{\perp}$,$\t I'_{\parallel}$)
and non-recurrent (CnR) where a (likely infinite) number of eigenvalues
vanish ($\t I_{\parallel}\leftrightarrow\t{II}_{\parallel}$).

- CR have a massive degenerate spectrum with non-zero imaginary parts,
therefore allowing for recurrent dynamics in the infinite-$s$ limit.
While the infinite-$s$ limit does not include a stable-steady state,
our variational approach, together with symmetry considerations, can
be used to predict both the magnetization and the entropy to leading
order in $s$. In this phase, we have that $\lim_{s\to\infty}\left\Vert \av{\bs S}\right\Vert /s<1$
and the von Neumann entropy diverges logarithmically with $s$.

- In CC phases a stable steady state may still exist. In this case
the degenerate spectrum coexists with a regular one that is well approximated,
as for NCS, by linearizing the Liouvillian around the stable fixed-point.
Moreover, the finite-$s$ steady-state are well approximated by those
obtained perturbatively from the linearized Liouvillian. This implies
that steady-state observables have a convergent $1/s$ expansion and
that the entropy of the steady-state is finite in the infinite-$s$
limit.

- For CnR systems, eigenvalues approach zero with a spectral gap that
vanishes as a power law. Here, the fitted numerical value is compatible
with a mean-field exponent $s^{-1/2}$. This, together with the perturbative
results obtained in region $\t I_{\parallel}$, suggests that the
approach to the steady-state for a generic observable, $\av{O\left(t\right)}-\av{O\left(\infty\right)}$,
follows a scaling function of the from $s^{-\frac{1}{2}}\Phi\left(\abs{\lambda}t,\abs{\lambda}^{2}s^{\frac{1}{2}}\right)$,
where $\lambda$ is the eigenvalue of the linearized problem that
vanishes at the transition. Assuming a scaling hypothesis, this implies
a $t^{-2}$ power law relaxation at the infinite-$s$ limit. However,
with the system sizes available to us, we were not able to numerically
confirm this prediction. For a CnR, the steady-state entropy is observed
to grow logarithmically with increasing $s$.

Although our classification focuses only on the properties of stable
and marginal steady-states, we have also shown that the low-lying
spectrum of the Liouvillian operator in the large $s$ limit \emph{cannot}
be reproduced only by analyzing the stable fixed points. Instead,
the spectrum is obtained as a superposition of two sets of eigenvalues,
coming from the stable and unstable fixed points. Since these eigenvalues
with a small real part rule the decay to the steady-state at large
times, the decay rates also carry information about the unstable fixed
points. Such understanding is relevant for experimental setups aimed
at studying the characteristic timescales described in Sec. \ref{subsec:Spectrum and dynamics}.

\section{Conclusion\label{sec:Discussion}}

In summary, we present a detailed analysis of the LMG model, featuring
a collective spin system, in the presence of a Markovian dissipative
environment. Motivated by recent prototypes of engineered atomic spin
devices we focus on two polarization cases. Our analysis is also of
interest to other variants of the dissipative LMG model that have
previously been studied in the contexts of quantum optics and cold
atomic setups. By employing a variational approach, as well as a $1/s$
perturbative method, we are able to systematically study the model.
Despite its apparent simplicity, this model exhibits a rich phase
diagram where different phases are shown to possess qualitatively
different steady-state and dynamical properties. We identify a number
of different phases and provide a tentative classification with terms
of their spectral and steady-state properties (see Tab. \ref{tab:Classification-of-steady-state}).

One of the open issues, not addressed in the present work, is to understand
the nature of the coexisting region near first order phase transitions.
Detailed studies \citep{dombi2013optical,mavrogordatos2017quantum}
have already reveled some of the properties of distribution functions
near the transition. However, in the coexisting region, a criterion
to predict which fixed point is realized at finite $s$, similar to
Maxwell's construction for equilibrium first order phase transitions
\citep{callen1998thermodynamics}, is still lacking.
\begin{acknowledgments}
We gratefully acknowledge discussions with S. Kirchner, A. Shakirov.
PR acknowledges support by FCT through the Investigador FCT contract
IF/00347/2014 and Grant No. UID/CTM/04540/2013.
\end{acknowledgments}

\appendix

\section{Equations of Motion in the Large $s$ Limit\label{sec:Equations-of-Motion}}

In this section, we present the details of a derivation of the semi-classical
equations of motion of the model. We do this in the next two sub-sections
in two slightly different ways. The first is the usual semi-classical
analysis. The second method consists of approximating the dynamics
by constraining the possible states within a family of variational
density matrices. To treat both parallel and perpendicular cases at
the same time, in this section, we assume that the Hamiltonian and
the jump operators are generically given by 
\begin{align}
H & =-\sum_{\alpha}\left(h^{\alpha}S_{\alpha}+\frac{1}{2s}\gamma^{\alpha}S_{\alpha}^{2}\right)\\
W_{i} & =\frac{1}{\sqrt{2s}}\sum_{\alpha}\eta_{i}^{\alpha}S_{\alpha}
\end{align}
where $h^{\alpha=x,y,z}$ and $\gamma^{\alpha}$ are real and $\eta_{i}^{\alpha}$
are complex parameters.

\subsection{Semi-classical dynamics \label{sec: Semiclassical dynamics}}

A close set of equations of motion in the semi-classical limit is
obtained assuming that, for a typical state, $\av{S_{\alpha}S_{\beta}}=\av{S_{\alpha}}\av{S_{\beta}}+O\left(s^{1}\right)$.
Assuming this factorization in the equations of motion for the magnetization
\begin{align}
\pd_{t}\av{S_{\beta}} & =\tr\left[S_{\beta}\L\left(\rho\right)\right]
\end{align}
one obtains the semi-classical equations of motion for the quantity
$n_{\alpha}=\frac{1}{s}\av{S_{\alpha}}$:
\begin{align}
\pd_{t}n_{\beta} & =\sum_{\alpha\gamma}\varepsilon_{\alpha\beta\gamma}h_{\alpha}n_{\gamma}+\sum_{\alpha\gamma}\varepsilon_{\alpha\beta\gamma}\gamma_{\alpha}n_{\alpha}n_{\gamma}\nonumber \\
 & -\sum_{i,\alpha\alpha\text{'}\gamma}\frac{1}{2}\varepsilon_{\alpha\beta\gamma}\im\left[\bar{\eta}_{i}^{\alpha'}\eta_{i}^{\gamma}\right]n_{\alpha}n_{\alpha}\label{eq:class}
\end{align}
where $\varepsilon_{\alpha\beta\gamma}$ is the anti-symmetric tensor.

The stability of the fixed-points of the semi-classical dynamics,
i.e. points obeying $\pd_{t}n_{\beta}=0$, is obtained by linearizing
the equations of motion in their vicinity
\begin{align*}
\pd_{t}\delta n_{\beta} & =M\left(n_{\beta}^{*}\right)\delta n_{\beta},
\end{align*}
where $n_{\beta}^{*}$ is the value of the fixed-point and $\delta n_{\beta}=n_{\beta}-n_{\beta}^{*}$.

Besides the trivial fixed point with $\abs{\bs n}=0$, which is found
to be generically unstable, all the other fixed points found have
$\abs{\bs n}=1$.

\subsection{\label{sec:Variational-density-matrix}Variational density matrix}

Here we detail the variational approach employed in the main text.
The results of this approach only differ from those in the previous
section for phase $0_{\perp}$ and $\text{I'}_{\perp}$ , where it
allows to find a line of variational steady-states to which the magnetization
vector of the finite-$s$ steady-state belongs.

The variational states are parameterized by:

\begin{align}
\rho\left(\boldsymbol{m}\right) & =\frac{e^{\bs m.\bs S/s}}{Z_{\bs m}}\label{eq:var_matrix}
\end{align}
with $Z_{\bs m}=\tr\left[e^{\bs m.\bs S/s}\right]$. This family of
states includes thermal states of Hamiltonian that are linear in $\bs S$.
Within this family, expectation values $\av{S_{\alpha}}$ and $\av{S_{\alpha}S_{\beta}}$
are given by

\begin{align}
\av{\bs S} & =\bs R.\av{\bs S}_{z}\\
\av{\bs S.\bs S^{T}} & =\bs R.\av{\bs S.\bs S^{T}}_{z}.\bs R^{T}
\end{align}
where $\av{...}_{z}=\tr\left[...e^{\abs{\bs m}S^{z}/s}\right]/\tr\left[e^{\abs{\bs m}S^{z}/s}\right]$
and $\bs R$ is a rotation matrix chosen such that $\bs m=\abs{\bs m}\bs R.\bs e_{z}$.
In the large $s$ limit these expressions simplify to 
\begin{align}
\av{\bs S}/s & =L\left(m\right)\frac{\bs m}{m}\\
\av{\bs S.\bs S^{T}}/s^{2} & =G(m)\bs m.\bs m^{T}+\frac{L(m)}{m}\bs 1
\end{align}
where
\begin{align}
L\left(x\right) & =\coth\left(x\right)-\frac{1}{x}\\
G(x) & =\frac{x^{2}-3x\coth(x)+3}{x^{4}}
\end{align}
Replacing this expressions in the equations of motion one obtains
\begin{align}
\pd_{t}\left[m_{\beta}\frac{L\left(m\right)}{m}\right] & =Y_{\beta}\label{eq:motion_inf}
\end{align}
with
\begin{align}
Y_{\beta} & =\sum_{\alpha\gamma}\epsilon_{\alpha\beta\alpha'}h_{\alpha}\frac{L(m)}{m}m_{\alpha'}+\sum_{\alpha\alpha\text{'}\gamma}\left[\frac{1}{2}\epsilon_{\alpha\beta\alpha'}\left(\gamma_{\alpha}-\gamma_{\alpha'}\right)\right.\nonumber \\
 & +\left.\frac{i}{4}\sum_{i}\left(\bar{\eta}_{i}^{\alpha'}\eta_{i}^{\gamma}\varepsilon_{\alpha\beta\gamma}-\bar{\eta}_{i}^{\gamma}\eta_{i}^{\alpha}\varepsilon_{\alpha'\beta\gamma}\right)\right]\times\nonumber \\
 & \left[G(m)m_{\alpha}m_{\alpha\text{'}}+\frac{L(m)}{m}\delta_{\alpha\alpha'}\right]
\end{align}
Steady-states must satisfy the condition $\sum_{\beta}Y_{\beta}m_{\beta}=0$.\textbf{
}For $Y_{\beta}\neq0$ this implies: $\abs{\bs m}\rightarrow\infty$
(fully polarized state) or $\sum_{i,\alpha\gamma}\varepsilon_{\alpha\beta\gamma}\left(\bar{\eta}_{i}^{\alpha}\eta_{i}^{\gamma}-\bar{\eta}_{i}^{\gamma}\eta_{i}^{\alpha}\right)=0$
for all $\beta$. Since the second condition is not verified in either
models, steady-states must be fully polarized and the equations for
steady-states for $\hat{\bs m}=\bs m/\abs{\bs m}$ reduce to those
of $\bs n$ in the last section, for $\abs{\bs n}=1$. Therefore,
for fully polarized steady-states both approaches coincide. We may
also have solutions satisfying $Y_{\beta}=0$. Although a general
analytical treatment of the phase diagram of these solutions is beyond
this paper's scope, we propose that the existence of these solutions
lead to the recurrent regions observed. In general, a solutions of
$Y_{\beta}=0$ will be a continuous line of marginal points connecting
the marginal (or saddle) steady-states obtained semi-classically.
In this paper, such marginal line only occurs for $\bs p\perp\bs h$
and in the plane $z=0$

\begin{figure*}[!t]
\centering{}\includegraphics[height=5.5cm]{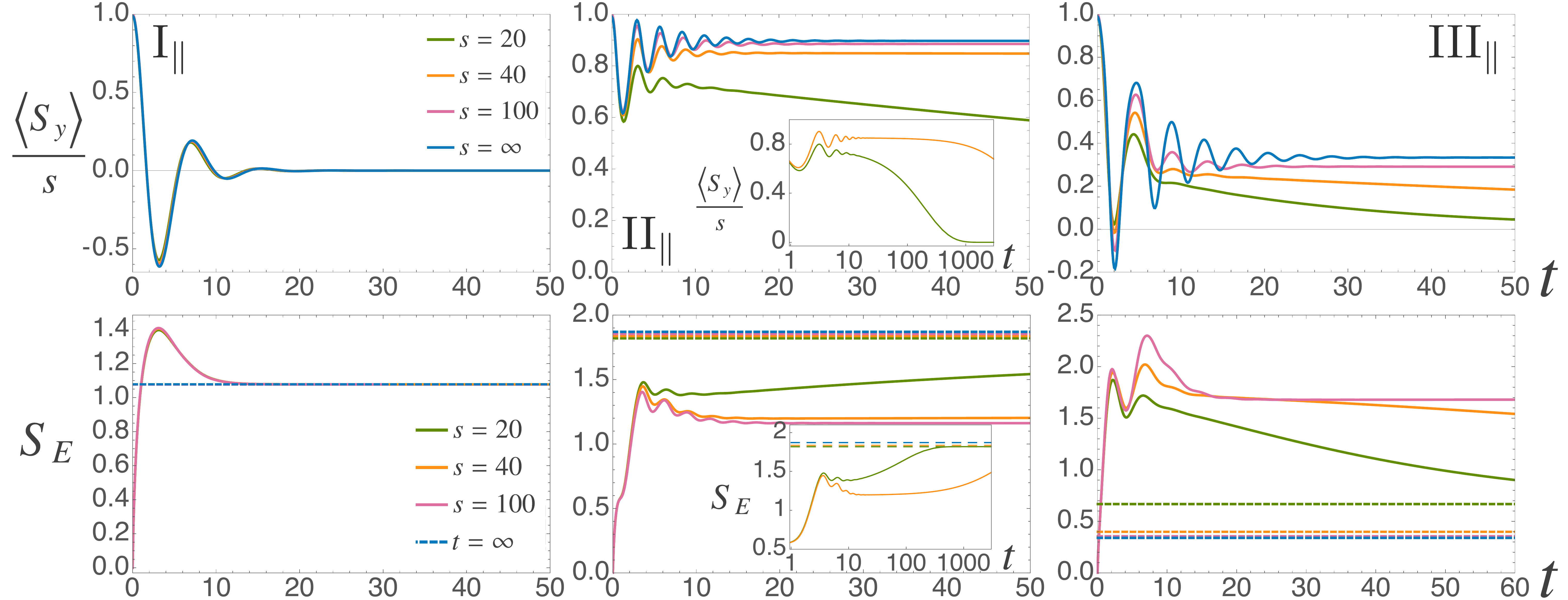}\caption{\label{fig:dyn_pz} Dynamics for $\protect\bs p\parallel\protect\bs h$:
Time evolution of $\protect\av{S_{y}}/s$ (upper panels) and von Neumann
entropy $S_{\text{E}}$ (lower panels) of an initial state polarized
along the $y$ direction, for different values of $s$. The parameters
are those of Fig.\ref{fig:pd_pz}.}
\end{figure*}
\begin{figure*}[!t]
\centering{}\includegraphics[height=5.5cm]{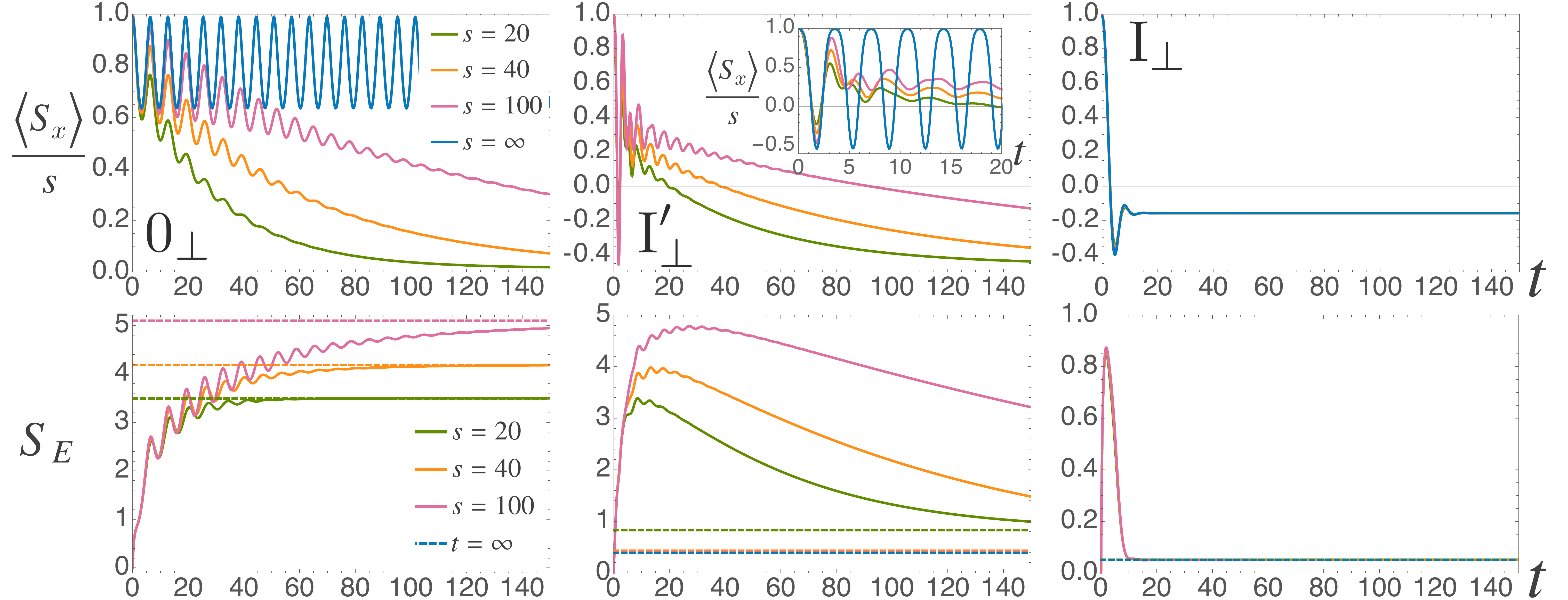}\caption{\label{fig:dyn_py} Dynamics for $\protect\bs p\perp\protect\bs h$
: Time evolution of $\protect\av{S_{x}}/s$ (upper panels) and von
Neumann entropy $S_{\text{E}}$ (lower panels) of an initial state
fully polarized along the $x$ direction for different values of $s$.
Parameters of Fig.\ref{fig:pd_pz}.}
\end{figure*}

\section{Dynamics for finite-s\label{sec:Dynamics-for-finite-s}}

In this section we present some helpful simulations of the magnetization
dynamics for finite-$s$ and all regions of the phase diagram. Figs.
\ref{fig:dyn_pz} and \ref{fig:dyn_py} show the long time dynamics
of a state initially polarized along the $y$ and $x$ direction,
respectively, for different spins ($s=20,40,100$).

In parallel case, Fig. \ref{fig:dyn_pz}, we highlight the visible
separation of timescales in region $\t{II}{}_{\parallel}$(center
panels), first a decay towards the mixed-symmetric state, followed
by an exponential decay towards the true steady-state. Unfortunately,
the same separation between the three timescales of region $\t{III}{}_{\parallel}$
is not so clear. The $s$ scaling convergence towards the infinite-$s$
magnetization dynamics (depicted as blue full line) and its entropy
(blue dashed lines) shows that the variational approach correctly
captures the dynamics in the large $s$ limit.

Similarly to the parallel case, the variational approach also captures
the dynamics in the large $s$ limit perpendicular case Fig. \ref{fig:dyn_py},
even when no stable steady-state exists. As discussed in Sec. \ref{subsec:Spectrum and dynamics}
the finite-$s$ steady-state in phase $\t I'_{\perp}$ is well approximated
by the unique stable steady-state at infinite-$s$ even though the
short time dynamics suggests a recurrent regime. In Fig. \ref{fig:dyn_py}-(center)
we plot the recurrent magnetization dynamics (upper plot) as blue
full line and the entropy of the stable steady-state as dashed blue
line (lower plot).

A similar situation occurs in region $\t 0{}_{\perp}$(Fig. \ref{fig:dyn_py}-left)
with the finite-$s$ steady-state being in the variational line with
$\av{S_{x}}=0$.

\section{The Linearized Lindblad Operator\label{sec:The Linearized Lindblad Operator}}

\subsection{Steady-state}

In this section we derive explicit expressions for the steady-state
of a linearized Lindblad operator. The presentation is done in a generic
way such that the approach can be used for more than one species of
bosons, in which case $A=\left\{ a_{1},a_{2},...,a_{n},a_{1}^{\dagger},a_{2}^{\dagger},...,a_{n}^{\dagger}\right\} ^{T}$.

As for the case of fermions \citep{Prosen2008a,Ribeiro2014e}, it
is useful to consider the single-body density matrix $\bs{\chi}=\av{A.A^{\dagger}}$.
The particular choice of the value of $\zeta$ and $\bar{w}_{0}$
in the Sec. \ref{sec:Linearized-Liouvillian-operator} leads to the
the vanish of the linear terms in $a$ and $a^{\dagger}$ , therefore
we consider that $H=A^{\dagger}.\bs H.A$, where single-particle Hamiltonian
is a $2n\times2n$ matrix respecting: $\bs H^{\dagger}=\bs H$ and
$\hat{\bs H}=\bs S\bs H^{T}\bs S$, with $\bs S=\left(\begin{array}{cc}
 & \bs 1\\
\bs 1
\end{array}\right)$, and $W_{i}=w^{\dagger}.A$, where $w$ is a $\mathbb{C}$-valued
vector with $2n$ components.

Under these assumptions the steady-state is Gaussian with a vanishing
first moment $\av A=0$. Thus, the second moment matrix $\bs{\chi}$
completely characterizes the steady state density matrix. This can
be seen explicitly for a density matrix of the form $\rho_{0}=e^{-\Omega_{0}}/Z_{0}$,
with $Z_{0}=\tr\left(e^{-\Omega_{0}}\right)$ and $\Omega_{0}=\frac{1}{2}A^{\dagger}.\bs{\Omega}_{0}.A$
where $\bs{\Omega}_{0}$ is Hermitian, $\bs{\Omega}_{0}^{\dagger}=\bs{\Omega}_{0}$,
and particle-hole symmetric, $\hat{\bs{\Omega}}_{0}=\bs S\bs{\Omega}_{0}^{T}\bs S$.
In which case the single-body density matrix is explicitly given by
\begin{align}
\bs{\chi}_{0} & =-n_{b}\left(-\bs J.\bs{\Omega}_{0}\right)\bs J
\end{align}
with $n_{b}\left(z\right)=\frac{1}{e^{z}-1}$ the Bose function and
$\bs J=\left(\begin{array}{cc}
\bs 1\\
 & -\bs 1
\end{array}\right)$.

Considering the adjoint of $\mathcal{L}$, $\mathcal{L}^{\t{ad}}=\mathcal{L}^{\dagger}$,
defined as $\tr\left[O\mathcal{L}\left(\rho\right)\right]=\tr\left[\mathcal{L}^{\t{ad}}\left(O\right)\rho\right]$,
for the linearized Lindblad operator the equation of motion $\pd_{t}A.A^{\dagger}=\mathcal{L}_{\text{lin}}^{\t{ad}}\left(A.A^{\dagger}\right)$
can be written as
\begin{align}
\pd_{t}A.A^{\dagger}= & -i\left[\bs K.A.A^{\dagger}-A.A^{\dagger}.\bs K^{\dagger}\right]+\bs J.\bs N.\bs J
\end{align}
where we defined 
\begin{align}
\bs K= & \bs J.\left(\bs H-i\bs{\Gamma}\right)\\
\bs N= & \sum_{\mu}w_{\mu}.w_{\mu}^{\dagger},
\end{align}
and 
\begin{align}
\bs{\Gamma}= & \frac{1}{2}\left(\bs N-\hat{\bs N}\right)\\
\bs M= & \frac{1}{2}\left(\bs N+\hat{\bs N}\right)
\end{align}
Taking the mean value with respect to some density matrix $\av{...}=\tr\left[...\rho\right]$,
we obtain the equation of motion for the single-body density matrix
given by 
\begin{align}
\pd_{t}\bs{\chi}= & -i\left[\bs K.\bs{\chi}-\bs{\chi}.\bs K^{\dagger}\right]+\bs J.\bs N.\bs J
\end{align}
A solution for the steady-state $\pd_{t}\bs{\chi}=0$ can be given
explicitly as 
\begin{align}
\bs{\chi}_{\infty}= & -i\sum_{\alpha\beta}\ket{\alpha}\frac{\bra{\tilde{\alpha}}\bs J.\bs N.\bs J\ket{\tilde{\beta}}}{\lambda_{\alpha}-\bar{\lambda}_{\beta}}\bra{\beta}
\end{align}
where $\ket{\alpha}$ and $\bra{\tilde{\alpha}}$, with $\braket{\tilde{\alpha}}{\beta}=\delta_{\alpha\beta}$,
are right and left eigenvectors of the operator $\bs K$ which can
be decomposed as $\bs K=\sum_{\alpha}\ket{\alpha}\lambda_{\alpha}\bra{\tilde{\alpha}}$.
It is worth noting that the particle-hole anti-symmetry $\bs K$,
i.e. $\hat{\bs K}\equiv\bs S\bs K^{T}\bs S=-\bs K^{\dagger}$, implies
that the eigenvectors of $\bs K$ appear in pairs: $\ket{\alpha}$
with eigenvalue $\lambda_{\alpha}$ and $\bs S\ket{\bar{\alpha}}$
with eigenvalue $-\bar{\lambda}_{\alpha}$.

Higher moments of $\rho_{0}$ can be completely determined by $\bs{\chi}_{0}$.
For example, the entanglement entropy is given by 
\begin{align*}
S & =\tr\left[\bs{\chi}_{\infty}\bs J\ln\left(\bs{\chi}_{\infty}\right)\right]
\end{align*}
for such quadratic bosonic model. This expression can be computed
from the eigenvalues of $\bs{\chi}_{0}\bs J$ (or of $\bs J\bs{\Omega}_{0}$)
that can be diagonalized by a symplectic transformation $\left(\bs J\bs U\right)\bs{\chi}_{0}\bs J\left(\bs J\bs U^{\dagger}\right)=\bs D\bs J$,
where $\bs D=\bs S\bs D\bs S$ is a diagonal matrix and $\bs U^{\dagger}\bs J\bs U=\bs J$.

\subsection{Spectrum and eigenstates of the Linearized Lindblad Operator}

In this section, we obtain the spectrum and eigenstates of the linearized
Lindblad operator by acting on the steady-state with a set of eigen-operators
of $\left[\mathcal{L}_{\text{lin}},...\right]$. We assume at first
that $\zeta$ and $\bar{w}_{0}$ are non-zero to see what are the
implications and set them to zero later. As for the last section,
the formalism is generic and can be used in the case there are several
species of bosons.

For the following treatment it is helpful to write the Lindblad operator
in the form

\begin{align}
\mathcal{L}_{\text{lin}} & =-\frac{i}{2}\bs{\mathfrak{a}}^{\dagger}\left[\begin{array}{cc}
\bs H-i\bs M & i\hat{\bs N}\\
i\bs N & -\bs H-i\bs M
\end{array}\right]\bs{\mathfrak{a}}+i\frac{1}{2}\tr\left(\bs K\right)
\end{align}
with 
\begin{align}
\bs{\mathfrak{a}} & =\left\{ a_{1}\otimes1,a_{2}\otimes1,...,a_{1}^{\dagger}\otimes1,...,1\otimes a_{1}^{T},...,1\otimes a_{1}^{\dagger T},...\right\} ^{T}
\end{align}
Since $\bs{\mathfrak{a}}.\bs{\mathfrak{a}}^{\dagger}-\left(\bs{\mathfrak{a}}^{\dagger T}.\bs{\mathfrak{a}}^{T}\right)^{T}=\bs{\mathfrak{J}}$,
with $\bs{\mathfrak{J}}=\t{diag}\left(\bs J,-\bs J\right)$, a transformation
$\mathfrak{a}\to\mathfrak{R}\mathfrak{a}$ that leaved the matrix
$\bs{\mathfrak{J}}$ invariant, i.e. $\mathfrak{R}^{\dagger}.\bs{\mathfrak{J}}.\mathfrak{R}=\bs{\mathfrak{J}}$,
respects the bosonic commutation relations.

In order to reveal the upper tridiagonal structure of $\mathcal{L}_{\text{lin}}$,
we perform the transformation $\tilde{\bs{\mathfrak{a}}}=\bs{\mathfrak{U}}\bs{\mathfrak{a}}$
with $\bs{\mathfrak{U}}=\frac{1}{\sqrt{2}}\left(\begin{array}{cc}
\bs 1 & \bs 1\\
\bs 1 & -\bs 1
\end{array}\right)$, yielding 
\begin{align}
\mathcal{L}_{\text{lin}} & =-\frac{i}{2}\tilde{\bs{\mathfrak{a}}}^{\dagger}\tilde{\bs{\mathfrak{J}}}\left[\begin{array}{cc}
\bs K & -2i\bs J\bs M\\
0 & \bs J\bs K^{\dagger}\bs J
\end{array}\right]\tilde{\bs{\mathfrak{a}}}+i\frac{1}{2}\tr\left(\bs K\right)\label{eq:L_lin_1}
\end{align}
where $\tilde{\bs{\mathfrak{J}}}=\bs{\mathfrak{U}}^{-1\dagger}\bs{\mathfrak{J}}\bs{\mathfrak{U}}^{-1}=\left[\begin{array}{cc}
0 & \bs J\\
\bs J & 0
\end{array}\right]$. Note that, in this basis, to preserve the bosonic commutation relations,
canonical transformations, $\tilde{\bs{\mathfrak{a}}}\to\mathfrak{\tilde{R}}\tilde{\bs{\mathfrak{a}}}$,
have to leave the form $\tilde{\bs{\mathfrak{J}}}$ invariant, i.e.
$\tilde{\mathfrak{R}}^{\dagger}.\tilde{\bs{\mathfrak{J}}}.\mathfrak{\tilde{R}}=\tilde{\bs{\mathfrak{J}}}$.
We can now use the upper tridiagonal from of Eq.(\ref{eq:L_lin_1})
find the transformation $\tilde{\bs{\mathfrak{a}}}=\mathfrak{\tilde{R}}\tilde{\bs{\mathfrak{b}}}$
with
\begin{align*}
\mathfrak{\tilde{R}} & =\left[\begin{array}{cc}
\bs R & \bs X\bs R^{-1\dagger}\bs J\\
0 & \bs J\bs R^{-1\dagger}\bs J
\end{array}\right]
\end{align*}
that diagonalizes $\mathcal{L}_{\text{lin}}$. Here the matrix $\bs R$
is taken to diagonalize $\bs K$, i.e. $\bs R^{-1}\bs K\bs R=\bs D$,
with $\bs D=\t{diag}\left(\lambda_{1},\lambda_{2},...,-\bar{\lambda}_{1},-\bar{\lambda}_{2},...\right)$
and $\bs X$, is defined by $\bs K\bs X-\bs X\bs K^{\dagger}=2i\bs J\bs M\bs J$,
and can be given explicitly as 
\begin{align*}
\bs X= & \sum_{\alpha\alpha'}2i\ket{\alpha}\frac{\bra{\tilde{\alpha}}\bs J\bs M\bs J\ket{\tilde{\alpha}'}}{\left(\lambda_{\alpha}-\bar{\lambda}_{\alpha'}\right)}\bra{\alpha'}
\end{align*}
In this basis we thus have
\begin{align*}
\mathcal{L}_{\text{lin}} & =-\frac{i}{2}\tilde{\bs{\mathfrak{b}}}^{\dagger}\tilde{\bs{\mathfrak{J}}}\left[\begin{array}{cc}
\bs D & 0\\
0 & \bs J\bs D^{\dagger}\bs J
\end{array}\right]\tilde{\bs{\mathfrak{b}}}+i\frac{1}{2}\tr\left(\bs D\right)
\end{align*}
 Finally transforming back $\bs{\mathfrak{b}}=\bs{\mathfrak{U}}^{-1}\tilde{\bs{\mathfrak{b}}}$
, defining the single mode variables $\bs{\mathfrak{b}}_{\alpha}=\left(\bs{\mathfrak{U}}^{-1}\mathfrak{\tilde{R}}^{-1}\bs{\mathfrak{U}}\bs{\mathfrak{a}}\right)_{\alpha}$
and the real and imaginary parts of the eigenvalues of $\bs K$, $\lambda_{\alpha}=\varepsilon_{\alpha}-i\gamma_{\alpha}$,
we can write 
\begin{align*}
\mathcal{L}_{\text{lin}} & =\sum_{\alpha}\left(-\frac{i}{2}\bs{\mathfrak{b}}_{\alpha}^{\dagger}\left[\begin{array}{cc}
\varepsilon_{\alpha}\bs 1 & -i\gamma_{\alpha}\bs J\\
i\gamma_{\alpha}\bs J & -\varepsilon_{\alpha}\bs 1
\end{array}\right]\bs{\mathfrak{b}}_{\alpha}+\gamma_{\alpha}\right)
\end{align*}
with $\bs{\mathfrak{b}}_{\alpha}=\left\{ b_{\alpha}\otimes1,b_{\alpha}^{\dagger}\otimes1,1\otimes b_{\alpha}^{T},1\otimes b_{\alpha}^{\dagger T}\right\} $,
where $b_{\alpha}$ are bosonic operators. In this form it is easy
to see that the eigen-operators of $\mathcal{L}_{\text{lin}}$ respecting
the property
\begin{align*}
\left[\mathcal{L}_{\text{lin}},\xi\right] & =\mu\xi
\end{align*}
with $\mu$ the respective eigenvector, are given by 
\begin{align*}
\xi_{\alpha\pm} & =\frac{1}{\sqrt{2}}\left(b_{\alpha}\otimes1\mp1\otimes b_{\alpha}^{T}\right)\\
\xi_{\alpha\pm}^{\dagger} & =\frac{1}{\sqrt{2}}\left(b_{\alpha}^{\dagger}\otimes1\mp1\otimes b_{\alpha}^{\dagger T}\right)
\end{align*}
with eigenvalues given respectively by 
\begin{align*}
\mu_{\alpha\pm} & =\mp\gamma_{\alpha}+i\varepsilon_{\alpha}\\
\bar{\mu}_{\alpha\pm} & =\mp\gamma_{\alpha}-i\varepsilon_{\alpha}
\end{align*}
There have the property $\left[\xi_{\alpha'\sigma},\xi_{\alpha\sigma'}^{\dagger}\right]=\delta_{\alpha\alpha'}\delta_{\sigma,-\sigma'}$
and $\left[\xi_{\alpha'\sigma},\xi_{\alpha\sigma'-}\right]=0$.

The eigen-operators, $\xi$, are useful because they allow to explicitly
construct the eigenstates of $\mathcal{L}_{\text{lin}}$ starting
from a reference state $\rho_{0}$, for which $\mathcal{L}_{\text{lin}}\left(\rho_{0}\right)=\Lambda_{0}\rho_{0}$
, for example 
\begin{align*}
\mathcal{L}_{\text{lin}}.\xi_{\alpha+}\left(\rho_{0}\right) & =\left(\mu_{\alpha+}+\Lambda_{0}\right)\xi_{\alpha+}\left(\rho_{0}\right),
\end{align*}
i.e. $\xi_{\alpha+}\left(\rho_{0}\right)$ is an eigenstate of $\mathcal{L}_{\text{lin}}$
with eigenvalue $\Lambda=\left(\mu_{\alpha+}+\Lambda_{0}\right)$.
In general we have
\begin{align*}
 & \mathcal{L}_{\text{lin}}\prod_{i}\xi_{\alpha_{i},\sigma_{i}}\prod_{j}\xi_{\alpha'_{i},\sigma'_{i}}^{\dagger}\left(\rho_{0}\right)=\\
 & \left(\sum_{i}\mu_{\alpha_{i},\sigma_{i}}+\sum_{j}\bar{\mu}_{\alpha'_{i},\sigma'_{i}}+\Lambda_{0}\right)\prod_{i}\xi_{\alpha_{i},\sigma_{i}}\prod_{j}\xi_{\alpha'_{i},\sigma'_{i}}\left(\rho_{0}\right)
\end{align*}
In the case where $\rho_{0}$ is the steady-state, i.e. $\Lambda_{0}=0$,
we have that, for a single mode $\alpha$, all the eigenstates of
$\mathcal{L}_{\text{lin}}$ can be written as $\rho_{n,m}=\left(\xi_{\alpha+}^{\dagger}\right)^{n}\left(\xi_{\alpha+}\right)^{m}\left(\rho_{0}\right)$
with eigenvalues $\Lambda_{n,m}=-\left(n+m\right)\gamma_{\alpha}-i\left(n-m\right)\varepsilon_{\alpha}$.
Moreover one can show that for the steady-state
\begin{align*}
\xi_{\alpha-}^{\dagger}\left(\rho_{0}\right)=\xi_{\alpha-}\left(\rho_{0}\right) & =0
\end{align*}

\subsection{Explicit example: Region $\text{I}_{\parallel}$ \label{subsec:Explicit-example:-Region}}

In most of the examples given in the main text, although linearization
can be simply performed, explicit expressions of physical quantities
are too cumbersome and bring no further significant understanding.
However it is instructive to present explicit results for a particular
case. In this section we illustrate the treatment of the preceding
sections for the particularly simple case of region $\t I_{\parallel}$
characterized by a stable and an unstable fixed points.

\subsubsection{Stable fixed-point}

Assuming $p>0$, region $\t I_{\parallel}$ is characterized by a
stable fixed point at $\alpha=0$, the linearized Lindblad operators
around this point is defined by the matrices 
\begin{align*}
\bs H_{\t s} & =\left(\begin{array}{cc}
-h-\frac{1}{2}\left(\gamma_{x}+\gamma_{y}\right) & \frac{1}{2}\left(\gamma_{y}-\gamma_{x}\right)\\
\frac{1}{2}\left(\gamma_{y}-\gamma_{x}\right) & -h-\frac{1}{2}\left(\gamma_{x}+\gamma_{y}\right)
\end{array}\right)
\end{align*}
and 
\begin{align*}
\bs N_{\t s} & =\left(\begin{array}{cc}
\frac{1}{2}(p+1)\Gamma & 0\\
0 & \frac{1}{2}(1-p)\Gamma
\end{array}\right)
\end{align*}
which yield eigenvalues of $\bs K_{s}$ given by $\lambda_{\t s,\pm}=\pm\sqrt{\left(h+\gamma_{x}\right)\left(h+\gamma_{y}\right)}-\frac{1}{2}i\Gamma p$
and to a single-particle density matrix given by
\begin{align*}
\bs{\chi}_{0}= & \left(\begin{array}{cc}
\kappa+1 & \bar{\delta}\\
\delta & \kappa
\end{array}\right)
\end{align*}
with
\begin{align*}
\kappa= & \eta\Big[\left(2h+\gamma_{x}+\gamma_{y}\right){}^{2}-4p\left(h+\gamma_{x}\right)\left(h+\gamma_{y}\right)\\
 & +(1-p)p^{2}\Gamma^{2}\Big]\\
\delta= & \eta\left(\gamma_{y}-\gamma_{x}\right)\left(2h+\gamma_{x}+\gamma_{y}+ip\Gamma\right)\\
\eta^{-1}= & 2p\left[4\left(h+\gamma_{x}\right)\left(h+\gamma_{y}\right)+\Gamma^{2}p^{2}\right]
\end{align*}
This expression yields a steady-state expectation value for the magnetization
that is given by

\begin{align*}
\av{\bs S} & =\left(-s+\kappa\right)\bs e_{z}
\end{align*}
and to the steady-state entanglement entropy 
\begin{align*}
S_{\t E}= & p_{+}\ln\left(p_{+}\right)+p_{-}\ln\left(-p_{-}\right)
\end{align*}
where $p_{\pm}=\frac{1}{2}\left(1\pm\sqrt{\left(1+2\kappa\right)^{2}-4\delta\text{\ensuremath{\bar{\delta}}}}\right)$
are the eigenvalues of $\bs{\chi}_{0}\bs J$.

In the main text, the expressions $\lambda_{\t s,\pm}$, $\av{\bs S}$
and $S_{\t E}$ are compared to the numerical results.

\subsubsection{Unstable fixed-point}

Although the unstable fixed point does not contribute to the steady-state
properties, its signatures can be traced in the spectrum. The linearized
Lindblad operator for $\alpha=\infty$, can most easily be obtained
considering the alternative Holstein-Primakoff (H-P) transformation
\begin{align}
S_{z} & =s-a^{\dagger}a\\
S_{-} & =a^{\dagger}\sqrt{2s-a^{\dagger}a}\\
S_{+} & =\sqrt{2s-a^{\dagger}a}\,a
\end{align}
After linearization we obtain 
\begin{align*}
\bs H_{\t{uns}} & =\left(\begin{array}{cc}
h-\frac{1}{2}\left(\gamma_{x}+\gamma_{y}\right) & \frac{1}{2}\left(\gamma_{y}-\gamma_{x}\right)\\
\frac{1}{2}\left(\gamma_{y}-\gamma_{x}\right) & h-\frac{1}{2}\left(\gamma_{x}+\gamma_{y}\right)
\end{array}\right)
\end{align*}
and 
\begin{align*}
\bs N_{\t{uns}} & =\left(\begin{array}{cc}
\frac{1}{2}(1-p)\Gamma & 0\\
0 & \frac{1}{2}(1+p)\Gamma
\end{array}\right)
\end{align*}
which gives $\lambda_{\t{uns},\pm}=\pm\sqrt{\left(h-\gamma_{x}\right)\left(h-\gamma_{y}\right)}+\frac{i\Gamma p}{2}$,
confirming that the point is indeed unstable for $p>0$. This fixed
point is responsible for a second ``cone'' of eigenvalues of $\mathcal{L}$,
determined by $\Lambda=in_{+}\lambda_{\t{uns},+}+in_{-}\lambda_{\t{uns},-}$
with $n_{\pm}=\mathbb{N}^{+}$and shifted by $-p\Gamma$.

\bibliographystyle{apsrev4-1}
\bibliography{Open_Lipkin}

\end{document}